\documentclass[final,authoryear,5p,times,twocolumn]{elsarticle}
\usepackage{amssymb,amsmath}
\usepackage{epsfig}
\usepackage{graphicx}

\begin{document}

\begin{frontmatter}

\title{Quasi-Gasdynamic Approach for Numerical Solution of Magnetohydrodynamic Equations}

\author{M.~V.~Popov\corref{cor1}}
\ead{mikhail.v.popov@gmail.com}

\author{T.~G.~Elizarova\corref{cor2}}
\ead{telizar@mail.ru}

\author{S.~D.~Ustyugov}

\address{Keldysh Institute of Applied Mathematics, 4, Miusskaya sq., Moscow, 125047, Russia}

\begin{abstract}
We introduce an application of the Quasi-Gasdynamic method for a solution of ideal magnetohydrodynamic equations in the modeling of compressible conductive gas flows. A  time-averaging procedure is applied for all physical parameters in order to obtain the quasi-gas-dynamic system of equations for magnetohydrodynamics. Evolution of all physical variables is presented in an unsplit divergence form. Divergence-free evolution of the magnetic field is provided by using a constrained transport method based on Stokes theorem. Accuracy and convergence of this method are verified on a large set of standard 1D and 2D test cases.
\end{abstract}

\begin{keyword}
magnetohydrodynamics\sep conservation laws\sep quasi-gasdynamics equations\sep finite-difference schemes
\end{keyword}

\end{frontmatter}


\section{Introduction}
\vspace{-2pt}
Up to now a variety of numerical methods to solve magnetohydrodynamic (MHD) equations is developed. Here we mention some of them, for example, Mac Cormack scheme~\citep{Yu}, Lax-Friedrichs scheme~\citep{Toth96}, the weighted essentially non-oscillatory scheme (WENO)~\citep{Jiang}, piecewise parabolic method (PPM)~\citep{Dai} and its local variant PPML~\citep{Pop08,Ust09}. All of them are equipped by limiter functions for suppressing oscillations near discontinuities and approximate Riemann solvers (e.g. Roe~\citep{Bri88}, HLLC~\citep{Li05} or HLLD~\citep{Miy05} solver). For practical applications a number of numerical codes have been developed up to now, for example for astrophycal simulations it should be mentioned Flash~\citep{FLASH}, Enzo~\citep{Enzo}, Athena~\citep{Athena} and Castro~\citep{Castro}.

An alternative  approach for numerical modeling of MHD equations ia presented below. It can be regarded as  generalization of regularized (or quasi-) gas dynamic (QGD) equation system  suggested in last three decades, e.g. ~\citep{Chet09,Elizarova,Sheretov}. 
 
The QGD equations are closely related to  the Navier-Stokes (NS) equation system and can be written in the form of the NS system except for additional strongly non-linear  terms. These additional terms contain the second-order space derivatives in a factor of a small parameter $\tau$ that has the dimension of time. These non-trivial  $\tau$- terms bring an additional non-negative entropy production that proves their dissipative character and contribute to the stability of the numerical solution, e.g.~\citep{Elizarova,Sheretov}. The $\tau$- terms depend on the solution itself and decrease in regions where space derivatives of the solution are small. In this sense the Quasi-Gasdynamic approach could be regarded as an approach with adaptive artificial viscosity.

Based on QGD equations,  a family of finite-difference homogeneous  schemes was constructed for numerical simulations of non-stationary gas flows. Efficiency, accuracy and simplicity of constructed algorithms are achieved due to regularization by  $\tau$-terms, included in all equations of the system. The QGD schemes are constructed by applying central-difference approximation of all space derivatives, but due to the chosen $\tau$ form they have the space accuracy of the first order. The QGD algorithm differs from the other first-order finite-volume methods, and among its advantages the validity of the entropy theorem must be mentioned. The last feature is a critical point for mathematical description and numerical methods in fluid dynamics.

The applicability of the QGD algorithm was demonstrated in~\citet{Elizarova2009}, where stability and accuracy  are studied numerically for ten classical Riemann test problems. The results revealed that the numerical solution monotonically converges to a self-similar one as the spatial grid is refined. Application of QGD algorithm for complex multidimensional gas-dynamical problems could be found in e.g.~\citet{Eliz01,Mate01,Gra04,Ch05}. The QGD algorithms are well suitable for unstructured computational grids and are quite naturally generalized  for the parallel realization using domain decomposition technique  in order to speed up computation, e.g.~\cite{Chet09}.

In this paper, the extension of the QGD method is presented for a solution of magnetohydrodynamic (QMHD) equations in the numerical modeling of compressible heat-conducting gas. The QMHD system is obtained based on averaging of initial MHD system over small time interval and taking into account viscosity and thermal conductivity. Averaging in time is done for all physical parameters, including magnetic field as well, contrary to works ~\citet{Sheretov}, where consideration of the magnetic field is taken mainly as an action of an external force. The first variants of the QMHD equations were obtained in 1997 together with calculations of electrically conductive liquid melt in external magnetic field, see citations in~\citep{Elizarova, Sheretov}.  As a result, correct computation is provided for complex cases of interaction with discontinuity of tangential component of the magnetic field. The QMHD algorithm was tested on a number of MHD problems  in one-dimensional and two-dimensional cases. Among them the 2D Riemann  problem for four states, blast wave propagation test, interaction of shock wave with a cloud and Orszag-Tang vortex problems are solved. The three last problems are also calculated in 3D formulation~\citep{Pop13} and will be presented in details in further publication. The first results for 1D and 2D MHD calculations were published in~\citet{prep1} and~\citet{prep2} consequently.

Generalization of QMHD system for non-ideal gas together with the entropy theorem with nonnegative dissipative function has been performed in ~\citep{Zlotnik}. Here the influence of external forces and heat source are also included.

In the first section, the development of the quasi-gas-dynamic system of magnetohydrodynamic equation is presented. QMHD equations, numerical algorithm and finite-difference scheme are described in sections~3-4. Solenoidal condition for magnetic field is presented in section~4. Extensive testing of the QMHD scheme for one-dimensional and two-dimensional cases is discussed in section~5. In conclusion we stated about the possibilities of the QMHD scheme for flow simulations.


\section{QMHD system of equations}
\vspace{-2pt}

The system of magnetohydrodynamic equations for viscous thermally conductive
gas can be written in the following form:
\begin{equation}
\label{adv_a}
\frac{\partial \rho}{\partial t}+\frac{\partial \rho u_j}{\partial x_j}=0,
\end{equation}
\begin{equation}
\label{adv1_a}
\frac{\partial \rho u_i}{\partial t}+\frac{\partial T_{ij}}{\partial x_j}=0,
\end{equation}

\begin{equation}
\label{adv3_a}
\frac{\partial E}{\partial t}+\frac{\partial Q_j}{\partial x_j}=0,
\end{equation}

\begin{equation}
\label{adv2_a}
\frac{\partial B_i}{\partial t}+\frac{\partial T^m_{ij}}{\partial x_j}=0,
\end{equation}
where

\begin{equation}
\label{adv4}
\nonumber
T_{ij}=\rho u_i u_j + p \delta_{ij} - \Pi_{ij} + \frac{1}{2} B^2\delta_{ij} - B_i B_j,
\end{equation}
\begin{equation}
\nonumber
\label{adv4}
Q_j=\rho u_j H - \Pi_{jk} u_k - q_j + u_j B^2 -B_j\left(u_k B_k\right),
\end{equation}

\begin{equation}
\label{adv4}
\nonumber
T^m_{ij}=u_j B_i - u_i B_j,
\end{equation}

\begin{equation}
\nonumber
\label{adv4}
E=\rho \varepsilon + \frac{\rho u^2}{2} + \frac{B^2}{2},\quad H=\frac{E+p}{\rho}
\end{equation}
\begin{equation}
\nonumber
\label{adv4}
u^2=u^2_x + u^2_y + u^2_z,\quad B^2=B^2_x + B^2_y + B^2_z.
\end{equation}

Here $\rho$ is the density, $u_i$ is the velocity component, $B_i$ is the component of
magnetic field, $E$ and $H$ are the total energy per unit volume, and total specific enthalpy
respectively, $p$ is the pressure, $\varepsilon$ is the specific internal energy. The system
of equations is supplemented by an equation of state, which in the case of ideal gas, has a form of
\begin{equation}
\nonumber
\label{adv}
p=(\gamma-1)\,\rho\,\varepsilon,
\end{equation}
where $\gamma$ is the adiabatic index.

Viscous stress tensor and heat-flux vector are defined as follows
\begin{equation}
\nonumber
\label{adv4}
\Pi_{ij}=\mu \left(\frac{\partial u_i}{\partial x_j}+\frac{\partial u_j}{\partial x_i}-\frac{2}{3}\delta_{ij}\frac{\partial u_k}{\partial x_k}\right),\\
\quad q_i=-k \frac{\partial T}{\partial x_i},
\end{equation}

where $\mu$ is the dynamic viscosity coefficient,
\begin{equation}
\label{klabel}
k=\frac{\mu \gamma R}{(\gamma-1)Pr}
\end{equation}
is the heat transfer coefficient, $Pr$ stands for the Prandtl number.


Implying the same approach as in ~\citet{Elizarova2011} we average the system~(\ref{adv_a})-(\ref{adv3_a}) over a small time interval $\delta t$ and calculate the time integrals approximately as
$$
\frac1{\delta t}\int_t^{t+\delta t}f(x_i,t')dt'\approx f(x_i,t)+\tau\frac{\partial f(x_i,t)}{\partial t},
$$
where $f$ denotes the averaging quantities. Here we assume that all include quantities have sufficient smoothness.  We assume that $\delta t$ is smaller than the characteristic hydrodynamic time and that
 all averaged values practically do not depend on time interval $\delta t$ in some range of $\delta t$~\citep{ Sheretov, Elizarova}. Parameter $\tau$ is small and related to interval of time averaging ($0\le\tau\le\delta t$), that is not strictly determined. So $\tau$ may be regarded as a free small parameter that will be precised  later.

Then the system~(\ref{adv_a})-(\ref{adv3_a}) could be rewritten as
\begin{equation}
\nonumber
\label{adv_b}
\frac{\partial}{\partial t}\left(\rho+\tau\frac{\partial\rho}{\partial t}\right)+\frac{\partial}{\partial x_j}\left(\rho u_j+\tau\frac{\partial\rho u_j}{\partial t}\right)=0,
\end{equation}
\begin{equation}
\nonumber
\label{adv1_b}
\frac{\partial}{\partial t}\left(\rho u_i+\tau\frac{\partial \rho u_i}{\partial t}\right)+\frac{\partial}{\partial x_j}\left(T_{ij}+\tau\frac{\partial T_{ij}}{\partial t}\right)=0,
\end{equation}
\begin{equation}
\nonumber
\label{adv3_b}
\frac{\partial}{\partial t}\left(E+\tau\frac{\partial E}{\partial t}\right)+\frac{\partial}{\partial x_j}\left(Q_j+\tau\frac{\partial Q_j}{\partial t}\right)=0,
\end{equation}

\begin{equation}
\nonumber
\label{adv2_b}
\frac{\partial}{\partial t}\left(B_i+\tau\frac{\partial B_i}{\partial t}\right)+\frac{\partial}{\partial x_j}\left(T^m_{ij}+\tau\frac{\partial T^m_{ij}}{\partial t}\right)=0.
\end{equation}

Now we drop the terms of the form $\displaystyle{\frac{\partial}{\partial t}\tau\frac{\partial}{\partial t}}$ in time
derivatives calculation supposing then they are small compared with the first-order time derivatives.

So, for example, the continuity equation writes  as
\begin{equation}
\nonumber
\label{adv}
\frac{\partial \rho}{\partial t}+\frac{\partial }{\partial x_i} \left(\rho u_i + \tau \frac{\partial }{\partial t} \rho u_i \right) = 0.
\end{equation}

Then we define the time derivative of momentum from the equation of motion
\begin{equation}
\nonumber
\label{adv1}
\frac{\partial \rho u_i}{\partial t}= -\frac{\partial }{\partial x_j} \left(\rho u_i u_j + p\delta_{ij} + \frac{1}{2} B^2\delta_{ij} - B_i B_j \right),
\end{equation}

where we restrict our consideration with first-order terms only, i.e. omit values of the orders
$ O(\tau\mu)$. 
Introducing the  definitions
\begin{equation}
\nonumber
\label{adv4}
w_i = \frac{\tau }{\rho } \frac{\partial }{\partial x_j} \left(\rho u_i u_j + p\delta_{ij} + \frac{1}{2} B^2\delta_{ij} - B_i B_j \right), \quad j_{mi}=\rho \left(u_i - w_i \right)
\end{equation}

we present the continuity equation in smoothed form as follows:
\begin{equation}
\label{adv_i1}
\frac{\partial \rho}{\partial t}+\frac{\partial j_{mi}}{\partial x_i}=0.
\end{equation}

In the same way, omitting the values of the order of $ O(\tau k)$  we write the other equations of QMHD system as:
\begin{equation}
\label{adv_i2}
\frac{\partial \rho u_i}{\partial t}+\frac{\partial T_{ij}}{\partial x_j}=\frac{\partial \Pi^n_{ij}}{\partial x_j},
\end{equation}
\begin{equation}
\label{adv_i4}
\frac{\partial E}{\partial t}+\frac{\partial F_i}{\partial x_i}+\frac{\partial Q^n_i}{\partial x_i}=\frac{\partial \Pi^n_{ij} u_j}{\partial x_i},
\end{equation}
\begin{equation}
\label{adv_i3}
\frac{\partial B_i}{\partial t}+\frac{\partial T^m_{ij}}{\partial x_j}=-\frac{\partial T^{mn}_{ij}}{\partial x_j},
\end{equation}
where
\begin{equation}
\nonumber
\label{adv4}
F_i=j_{mi}\left(H+\frac{B^2}{2\rho}\right) -B_i\left(u_k B_k\right),
\end{equation}
\begin{equation}
\nonumber
\label{adv4}
T_{ij}=j_{mi} u_j + p\delta_{ij} + \frac{1}{2} B^2\delta_{ij} - B_i B_j,
\end{equation}
\begin{equation}
\nonumber
\label{adv4}
\Pi^n_{ij}=\Pi_{ij} - \rho u_i \Delta u_j - \Delta p \delta_{ij} - \frac{1}{2} \Delta B^2 \delta_{ij} + \Delta \left(B_i B_j \right),
\end{equation}
\begin{equation}
\nonumber
\label{adv4}
Q^n_i=q_i + \rho u_i \Delta \varepsilon + \rho u_i \left(p+B^2\right) \Delta \frac{1}{\rho} \\
+ u_i\left(B_k \Delta B_k\right) - B_i\left(B_k \Delta u_k\right),
\end{equation}
\begin{equation}
\nonumber
\label{adv4}
T^{mn}_{ij}=\Delta u_j B_i - \Delta u_i B_j +  u_j \Delta B_i - u_i \Delta B_j,
\end{equation}
with $\tau$-terms presented using the definition $\tau \partial f/\partial t=\Delta f$. Here $\Delta$-terms are defined from the known magnetohydrodynamic equations 
\begin{equation}
\nonumber
\label{adv4}
\Delta \frac{1}{\rho} = - \tau \left(u_i \frac{\partial}{\partial x_i} \frac{1}{\rho} - \frac{1}{\rho} \\
\frac{\partial u_i}{\partial x_i}\right),
\end{equation}
\begin{equation}
\nonumber
\label{adv4}
\Delta u_i = - \tau \left(u_j \frac{\partial u_i}{\partial x_j} + \frac{1}{\rho} \frac{\partial p}{\partial x_i} \\
 + \frac{1}{\rho} \frac{\partial}{\partial x_j} \frac{B^2}{2} \delta_{ij} \\
 - \frac{1}{\rho} \frac{\partial B_i B_j}{\partial x_j} \right),
\end{equation}
\begin{equation}
\nonumber
\label{adv4}
\Delta \varepsilon = - \tau \left(u_i \frac{\partial \varepsilon}{\partial x_i} + \frac{p}{\rho} \\
\frac{\partial u_i}{\partial x_i}\right),
\end{equation}
\begin{equation}
\nonumber
\label{adv4}
\Delta p = - \tau \left(u_i \frac{\partial p}{\partial x_i} + \gamma p \\
\frac{\partial u_i}{\partial x_i}\right),
\end{equation}
\begin{equation}
\nonumber
\label{adv4}
\Delta B_i =\tau \frac{\partial}{\partial x_j} \left(u_i B_j - u_j B_i\right).
\end{equation}
\vspace{-6pt}

The system~(\ref{adv_i1})-(\ref{adv_i4}) is the approximation of the initial system~(\ref{adv_a})-(\ref{adv3_a}) with the order of $\tau$ and for $\tau =0$ equation system ~(\ref{adv_i1})-(\ref{adv_i4})  reduces to the classical system ~(\ref{adv_a})-(\ref{adv3_a}). Notice, that the averaging procedure and the introducing of the $\tau$-terms do not disturb the solenoidal condition for the magnetic field ${\rm div}{\bf B} = 0$.

\section{Numerical algorithm}
\vspace{-2pt}

Let us introduce a uniform mesh for coordinates $x,y$ with steps $h_x,h_y$, and time interval with step $\Delta t$. Values of all physical quantities, namely, the density, the velocity, the energy and the magnetic field are defined at the mesh nodes (cell centers). The flux values of all quantities are defined on the cell edges with semi-integer indices.

An explicit time-difference scheme of the following form is used to solve the derived system~(\ref{adv_i1})-(\ref{adv_i4}). For the density:
\begin{equation}
\nonumber
\label{adv4}
\overline \rho_{ij} = \rho_{ij} - \frac {\Delta t}{h_x}\left(j_{mi+1/2j}-j_{mi-1/2j}\right) - \\
\frac {\Delta t}{h_y}\left(j_{mij+1/2}-j_{mij-1/2}\right),
\end{equation}
for the momentum components:
\begin{multline*}
\overline {\rho u}_{k,ij} = \rho u_{k,ij} -\\ \frac {\Delta t}{h_x}\left(T^n_{k1,i+1/2j}-T^n_{k1,i-1/2j}\right) - \frac {\Delta t}{h_y}\left(T^n_{k2,ij+1/2}-T^n_{k2,ij-1/2}\right)+\\
\frac {\Delta t}{h_x}\left(\Pi^n_{k1,i+1/2j}-\Pi^n_{k1,i-1/2j}\right) + \frac {\Delta t}{h_y}\left(\Pi^n_{k2,ij+1/2}-\Pi^n_{k2,ij-1/2}\right),
\end{multline*}
for the total energy:
\begin{multline*}
\overline E_{ij}\!=E_{ij} - \frac {\Delta t}{h_x}\!\left(F_{1,i+1/2j}\!-F_{1,i-1/2j}\right)-\frac{\Delta t}{h_y}\!\left(F_{2,ij+1/2}\!-F_{2,ij-1/2}\right)-\\
 \frac {\Delta t}{h_x}\left(Q_{1,i+1/2j}-Q_{1,i-1/2j}\right) - \frac {\Delta t}{h_y}\left(Q_{2,ij+1/2}-Q_{2,ij-1/2}\right)+\\
\frac {\Delta t}{h_x}\!\left(\Pi^n_{1k}u_{k,i+1/2j}\!-\Pi^n_{1k}u_{k,i-1/2j}\right) + \frac {\Delta t}{h_y}\!\left(\Pi^n_{2k}u_{k,ij+1/2}\!-\Pi^n_{2k}u_{k,ij-1/2}\right).
\end{multline*}
To provide non-divergence of magnetic field we consider (\ref{adv_i3}) in the form of Stokes' theorem. The numerical algorithm to calculate the components of the magnetic field will be described in the next section.

The quantities with overbars here stand for the unknown values on a new time layer, which are computed via the known values on the previous time layer and the differences between the fluxes through the edges of the cell $(i,j)$. All quantities on the edges between cells are defined with simple averaging by adjacent cells, e.g. for the density as example it is
\begin{equation}
\nonumber
\label{adv}
\rho_{i+1/2}=0.5\left(\rho_i+\rho_{i+1}\right).
\end{equation}

We solve the system~(\ref{adv_i1})-(\ref{adv_i4}) in Euler approximation where the physical thermal conductivity and physical viscosity are neglected. All dissipative terms, containing $\mu$, $k$ and $\tau$ coefficients, are regarded as artificial regularization factors. The relaxation parameter and coefficients of the viscosity are linked and calculated as
\begin{equation}
\label{tau1}  
\tau = \alpha \frac {h}{c_f},\quad \mu = \tau \cdot p \cdot Sc,
\end{equation}
where $\alpha$ is the numerical coefficient to be taken in the range of $0.1-0.5$, $h=0.5\left(h_x+h_y\right)$. The Prandtl number and the Schmidt number could be set equal to one. The thermal conductivity is determined by~(\ref{klabel}). Coefficient $\alpha$ is chosen according with accuracy and stability of the numerical algorithm.

Note, that if we replace the step size $h$ by mean free path $\lambda$, then the expressions~(\ref{tau1})  with $\alpha =1$ exactly equal to expressions for  a mean intercollisional time and mean free path in  rarefied gases, obtained in kinetical theory, e.g., \cite{Bird}.

Time step $\Delta t$ is determined by the Courant condition
\begin{equation}
\nonumber
\label{adv}
\Delta t= \beta \cdot min \left(\frac {h_x}{max_{ij}\left(|u_{x,ij}|+c_{fx,ij}\right)},\frac {h_y}{max_{ij}\left(|u_{y,ij}|+c_{fy,ij}\right)}\right),
\end{equation}
where $\beta$ is the numerical Courant coefficient and in computations it falls in the range of $0.1-0.3$
in most cases, $c_f$ is the fast magnetosonic speed~\citep{Gardiner}.


In the above numerical scheme  all space derivatives are approximated by central differences, that provide the accuracy of order $O(h^2)$. The definition of $\tau$ (\ref{tau1}) decreases the accuracy down to $O(h)$. Formally, omitting the dependence $\tau\sim h$ and using high order central difference approximations one can increase the order of accuracy of the QMHD scheme. But resulting schemes will need special turning for $\tau$ value depending on the problem under consideration. For practical applications it is not convenient. Theoretical investigations and practical experience show, that  for $\tau$ in form (\ref{tau1}) with base values $\alpha=0.5$, $Sc=Pr=1$, QGD and QMHD algorithms uniformly describe a large variety  of gas dynamic flows. In special cases tuning of numerical coefficients $\alpha$, $Sc$ and $Pr$ improves a numerical solution compared with base results (e.g. for Riemann problems in~\citet{Elizarova2009}, for MHD flows see test~5.4 below).


The QGD numerical scheme for non-uniform space grids could be constructed by replacing $h_x$ and $h_y$ for $h_{x,\,i}$ and $h_{y,\,j}$. Here the uniform space grids are presented for simplicity reason.

The presented numerical algorithm is simple because it is explicit, only central-difference approximations of the derivatives are used and additional monotonization procedures as limiting functions  are not required.

\vspace{-6pt}

\section{Solenoidal condition for magnetic field}
\vspace{-2pt}

In the numerical solution of the QMHD system of equations, it is necessary to satisfy the solenoidal condition for the magnetic field~\citep{Toth}. For this purpose, Stokes' theorem is used
\begin{equation}
\nonumber
\label{es43}
\frac{\partial{\bf\,B}}{\partial\,t}=-\nabla\times{\bf E},
\end{equation}

where the electric field is derived from formula $\bf E =-\bf u \times \bf B$, according
to the MHD approximation. Components of the electric field are the corresponding components of the fluxes calculated through the edges of cells. Averaging these fluxes on adjacent cells, components of the electric field, defined on the edges of cells and required for the application of Stokes' theorem, can be obtained. The method of this type was also implemented in e.g.~\citet{Pop08,Ust09}.

In two-dimensional case, z-component of the electric field is required, determined at the nodes of the mesh as
\begin{multline*}
\nonumber
\label{es43}
E_{z,i+1/2j+1/2}=\frac {1}{4}\left(T_{21,i+1/2j}+T_{21,i+1/2j+1}\right.+\\ \left.T_{12,ij+1/2}+T_{12,i+1j+1/2}\right),
\end{multline*}

where $T_{ij}=T^m_{ij}+T^{mn}_{ij}$ includes $\tau$ regularizations. Using Stokes' theorem, components of the magnetic field for
the next time step are calculated as
\begin{equation}
\nonumber
\label{adv4}
B^{n+1}_{x,i+1/2j} = B^n_{x,i+1/2j} - \frac {\Delta t}{h_y}\left(E_{z,i+1/2j+1/2}-E_{z,i+1/2j-1/2}\right),
\end{equation}
\begin{equation}
\nonumber
\label{adv4}
B^{n+1}_{y,ij+1/2} = B^n_{y,ij+1/2} + \frac {\Delta t}{h_x}\left(E_{z,i+1/2j+1/2}-E_{z,i-1/2j+1/2}\right).
\end{equation}

Components of the magnetic field at the mesh center $\left(i,j\right)$  are obtained by half of the sum
\begin{equation}
\nonumber
\label{es43}
B_{x,ij}=\frac {1}{2}\left(B_{x,i+1/2j}+B_{x,i-1/2j}\right),
\end{equation}
\begin{equation}
\nonumber
\label{es43}
B_{y,ij}=\frac {1}{2}\left(B_{y,ij+1/2}+B_{y,ij-1/2}\right).
\end{equation}

Let us remark that for the solution of the QMHD system for three-dimensional case all three components of the magnetic field should be redefined.

\vspace{-6pt}

\section{Numerical tests}
\vspace{-2pt}

Described numerical scheme was verified on several distinctive one-dimensional and two-dimensional
MHD problems to check its convergence and accuracy for Euler equations.  For all tests, uniform mesh with constant step in each direction was set and equation of state of ideal gas was used. In one-dimensional case, all computations were performed on the interval $x\in[0\ldots1]$. Initial values of vector $\bf V$ components, representing physical parameters, were defined on left and right sides from the middle
point of the interval as following:
\begin{equation}
\nonumber
\label{es13}
{\bf V}=\left\{
 \begin{aligned}
  {\bf V}^L, \quad \mbox{if}\ x \le 0.5\,,\\
  {\bf V}^R, \quad \mbox{if}\ x > 0.5\,.\\
 \end{aligned}
\right.
\end{equation}
The number of numerical cells was equal to $N$, the final time for computations was notated as $T$.
Boundary conditions were matched with corresponding initial conditions at the limits of the computational
region. For every value $N$, we calculate relative error $\delta_N$ and the real order of accuracy $R_N$, using the following definitions:
\begin{equation}
\nonumber
\label{eqxx9}
\delta_N=\frac {1}{8}\sum^{i=8}_{i=1} E_N\left(V_i\right),
\end{equation}
\begin{equation}
\nonumber
\label{eqxx9}
E_N\left(V_i\right) =\frac {\sum^{i=8}_{i=1} |V_{i,k}-V^{ex}_{i,k}|}{\sum^{i=8}_{i=1} |V^{ex}_{i,k}|},
\end{equation}
\begin{equation}
\nonumber
\label{eqxx9}
R_N=log_{2}\left(\delta_{N/2}/\delta_N\right).
\end{equation}

\subsection{Riemann problem with initial discontinuity of transversal component of magnetic field}

Initial conditions are given by~\citet{Dai}:
\begin{equation}
\nonumber
\left(\rho^{\,L},u^{\,L},v^{\,L},w^{\,L},B_y^{\,L},B_z^{\,L},p^{\,L}\right)=(1.0,0,0,0,1.0,0,1.0),
\end{equation}
\begin{equation}
\nonumber
\left(\rho^{\,R},u^{\,R},v^{\,R},w^{\,R},B_y^{\,R},B_z^{\,R},p^{\,R}\right)=(0.125,0,0,0,-1.0,0,0.1).
\end{equation}

\begin{figure}[!h]
\centering
\includegraphics[width=0.48\textwidth]{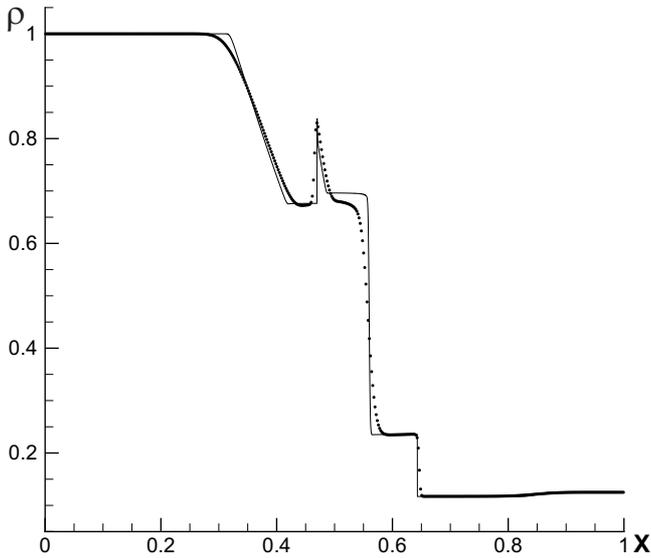}
\caption{Test 5.1. Density distribution obtained on the grid $N=512$ in comparison with the exact one.\label{Test1_D}}
\end{figure}

\begin{figure}[!h]
\centering
\includegraphics[width=0.48\textwidth]{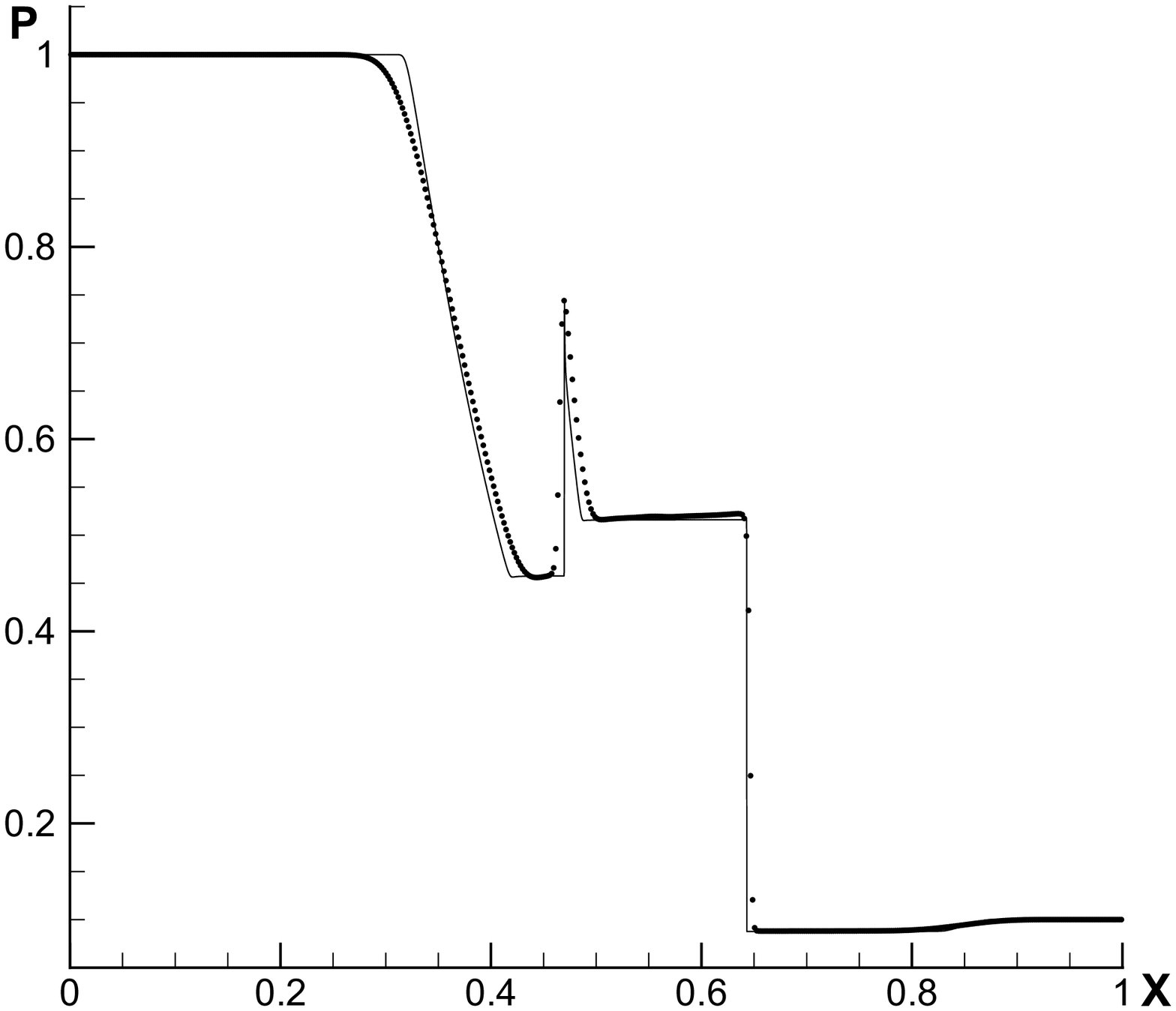}
\caption{Test 5.1. The same for the pressure.
\label{Test1_P}}
\end{figure}

\begin{figure}[!h]
\centering
\includegraphics[width=0.48\textwidth]{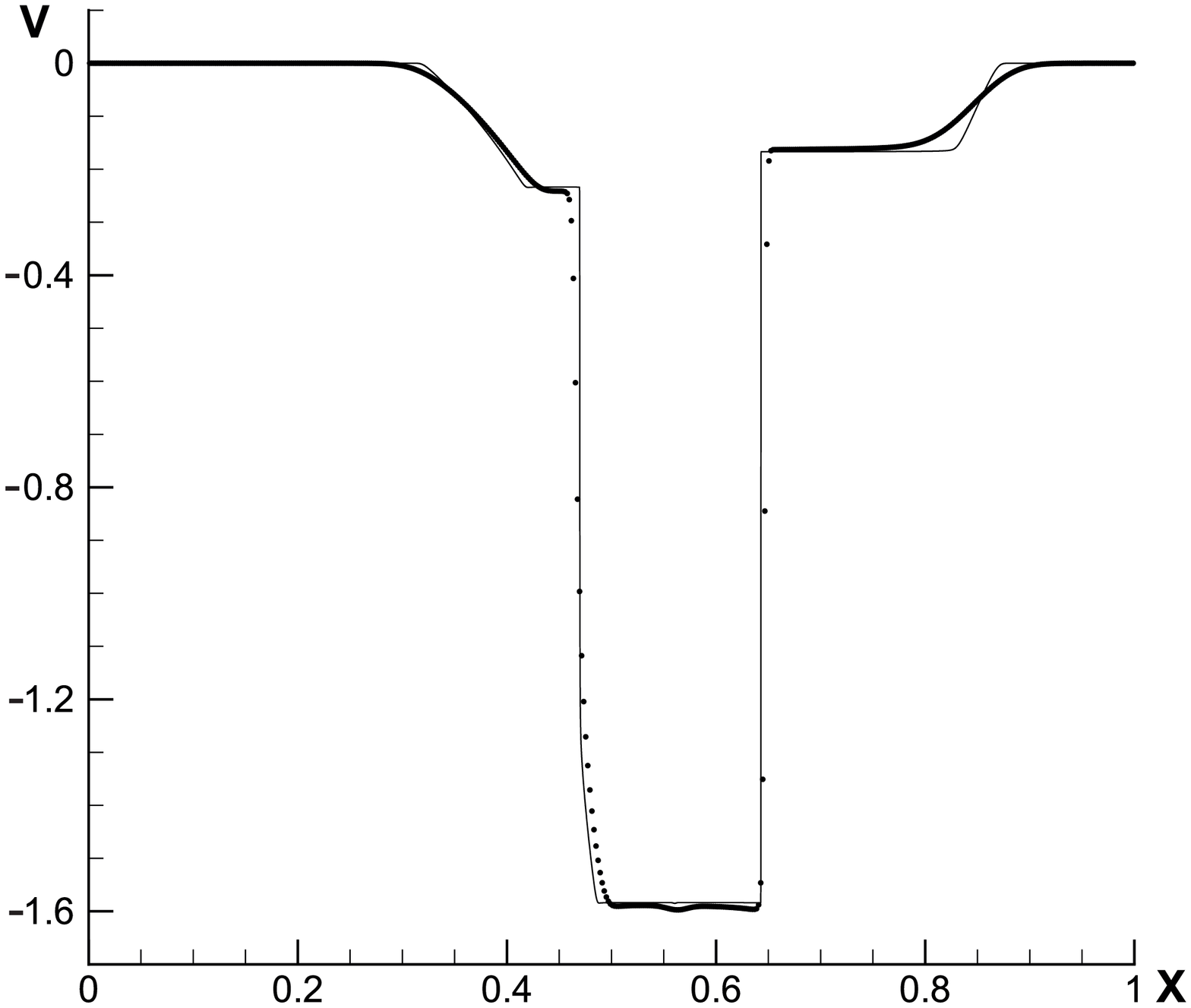}
\caption{Test 5.1. The same for $y$-component of the velocity.
\label{Test1_Vy}}
\end{figure}

\begin{figure}[!h]
\centering
\includegraphics[width=0.48\textwidth]{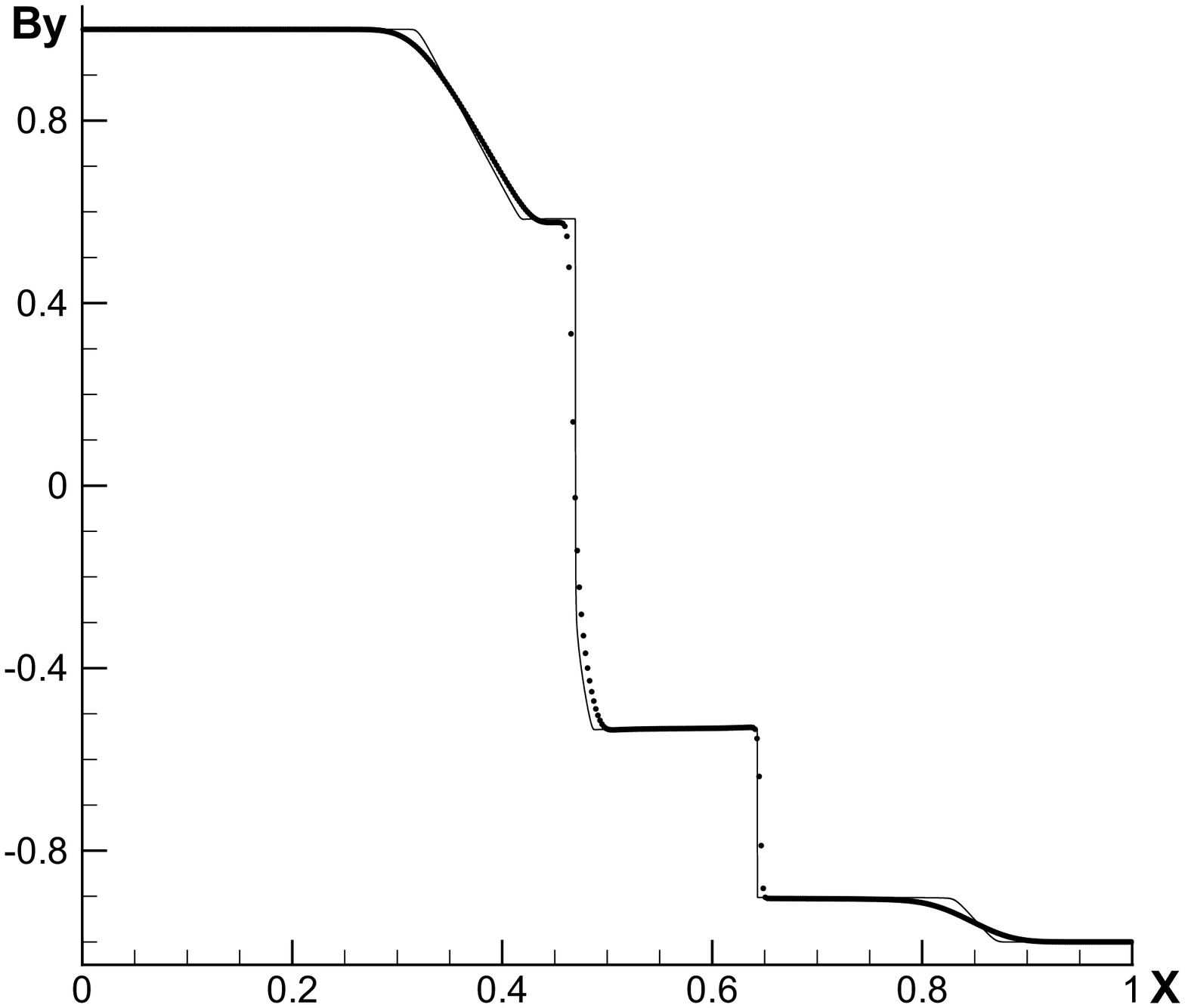}
\caption{Test 5.1. The same for $y$-component of the magnetic field.
\label{Test1_By}}
\end{figure}

Component of the magnetic field $B_x=0.75$, adiabatic index $\gamma=2$, $N=512$, $T=0.1$. The computational parameters are: Courant number $\beta = 0.2$, regularization coefficient $\alpha = 0.4$. Computation results are presented on Fig.~\ref{Test1_D}-\ref{Test1_By}. In this problem, the solution of the QMHD system of equations consists of fast rarefaction wave, moving to the left, intermediate shock wave and slow rarefaction wave, contact discontinuity, slow shock wave and one more fast rarefaction wave, moving to the right. The detailed discussion of this solution can be found in~\citet{Jiang}. On the figures, approximate solution is depicted by dots and exact solution is shown as solid line (it was obtained on the grid with $N=20000$ by the same code). Numerical scheme of QMHD system accurately represents all physical discontinuities and distribution behavior of all quantities without visible oscillations.

Notice, that similar pick in density and pressure distribution, as we see in Figs.~\ref{Test1_D} and.~\ref{Test1_P}, are presented in many other computations of this problems performed by high order methods with limiters, see, e.g.~\citet{Jiang12,Pop08}.

\begin{table}[!h]
\caption{Test 5.1. The average relative errors and the rates of convergence at $T=0.1$.}
\begin{center} \footnotesize
\begin{tabular}{|c|c|c|}
\hline
\rule{0pt}{12pt}\rule{6pt}{0pt}$N$\rule{6pt}{0pt} &\rule{25pt}{0pt} $\delta_N$\rule{25pt}{0pt} &\rule{8pt}{0pt} $R_N$\rule{8pt}{0pt}\\
\hline
128  & $6.91\times10^{-2}$ & -    \\
256 & $4.34\times10^{-2}$ & 0.67\\
512 & $2.74\times10^{-2}$ & 0.67\\
1024 & $1.52\times10^{-2}$ & 0.85\\
\hline
\end{tabular}
\end{center}
\label{tbl1}
\end{table}

As can be seen from Table~\ref{tbl1}, with mesh refinement the scheme error decreases with speed typical to first-order schemes. The quality of the numerical solution could be improved by adjustment of the numerical coefficients $Pr$ and $Sc$ (\ref{tau1}) for this test case.

The same test was performed in~\citet{Pop08} using the PPML method that is third order-accurate in space and second order-accurate in time. The PPML results obtained on the grid with 512 cells are very similar to the presented above except the region of contact discontinuity. The PPML resolves contact discontinuities better but it requires more computational cost than QMHD. Still QMHD requires more detailed computational grid to obtain the comparative quality of a solution.

\subsection{Riemann problem with formation of all forms of discontinuities}

Here the solution consists of two fast shock waves with the speed equal to 1.84 and 1.28 of
Mach number and directed to the left and to the right respectively, two slow shock waves,
moving to the left and to the right with the speed 1.38 and 1.49 of Mach number correspondingly, one rotational and two contact discontinuities. Initial conditions are given by~\citet{Dai}:
\begin{multline*}
\left(\rho^{\,L},u^{\,L},v^{\,L},w^{\,L},B_y^{\,L},B_z^{\,L},p^{\,L}\right)=\left(0.18405, 3.8964, 0.5361,\right.\\\left.2.4866, 2.394/\sqrt{4\pi}, 1.197/\sqrt{4\pi}, 0.3641\right),
\end{multline*}
\begin{multline*}
\left(\rho^{\,R},u^{\,R},v^{\,R},w^{\,R},B_y^{\,R},B_z^{\,R},p^{\,R}\right)=\\
\left(0.1,-5.5,0,0,2/\sqrt{4\pi},1/\sqrt{4\pi},0.1\right).
\end{multline*}

\begin{figure}[!t]
\centering
\includegraphics[width=0.48\textwidth]{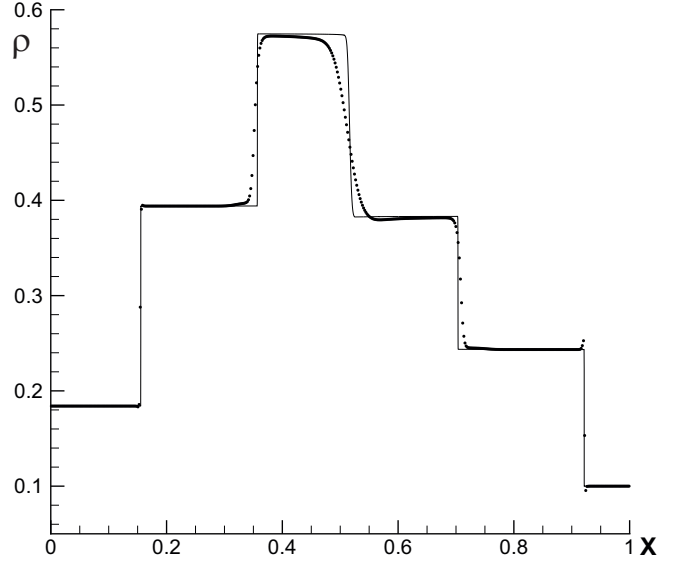}
\caption{Test 5.2. Density distribution obtained on the grid $N=512$ in comparison with the exact one.
\label{Test3_D}}
\end{figure}

\begin{figure}[!h]
\centering
\includegraphics[width=0.48\textwidth]{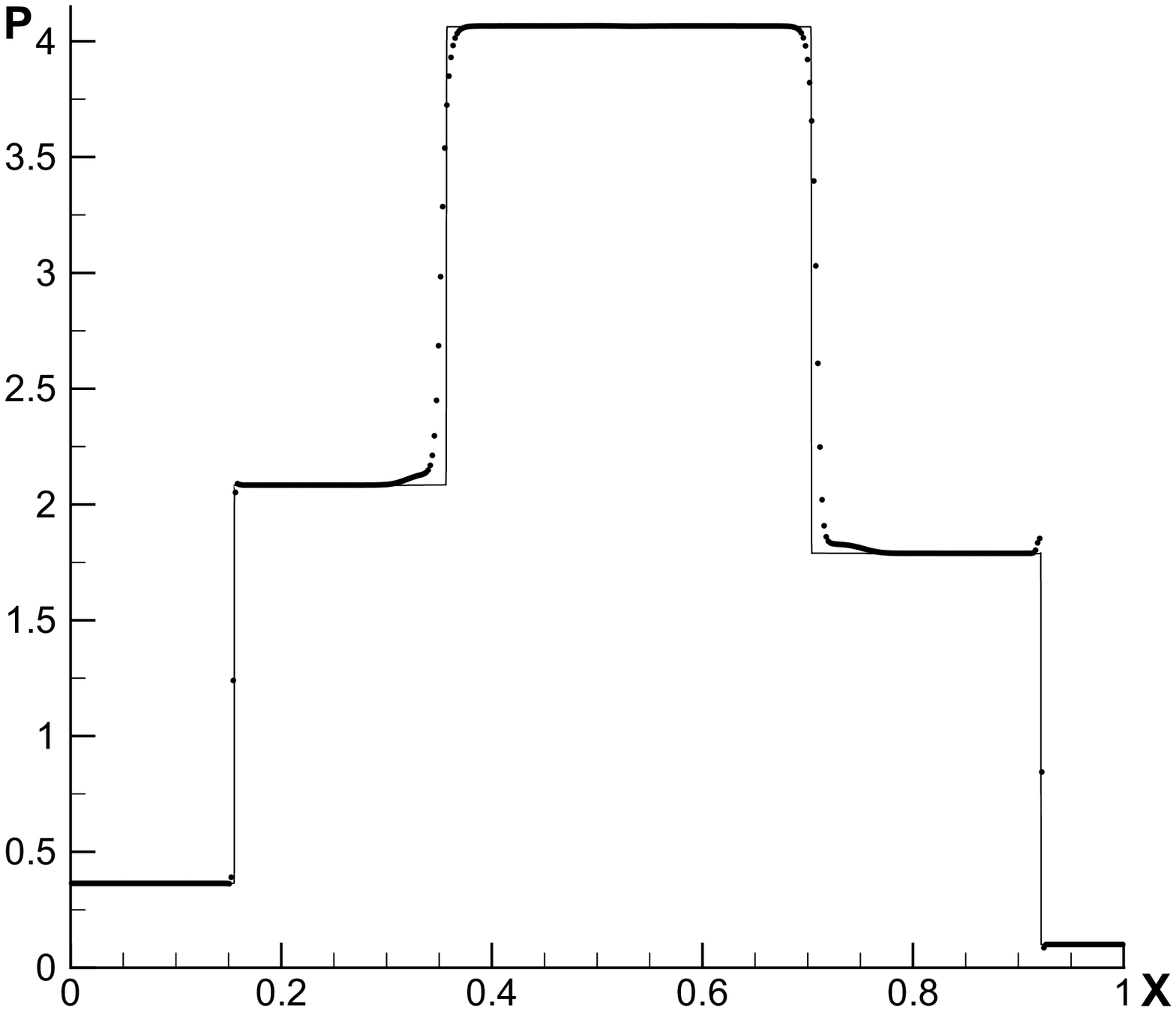}
\caption{Test 5.2. The same for the pressure.
\label{Test3_P}}
\end{figure}

\begin{figure}[!h]
\centering
\includegraphics[width=0.48\textwidth]{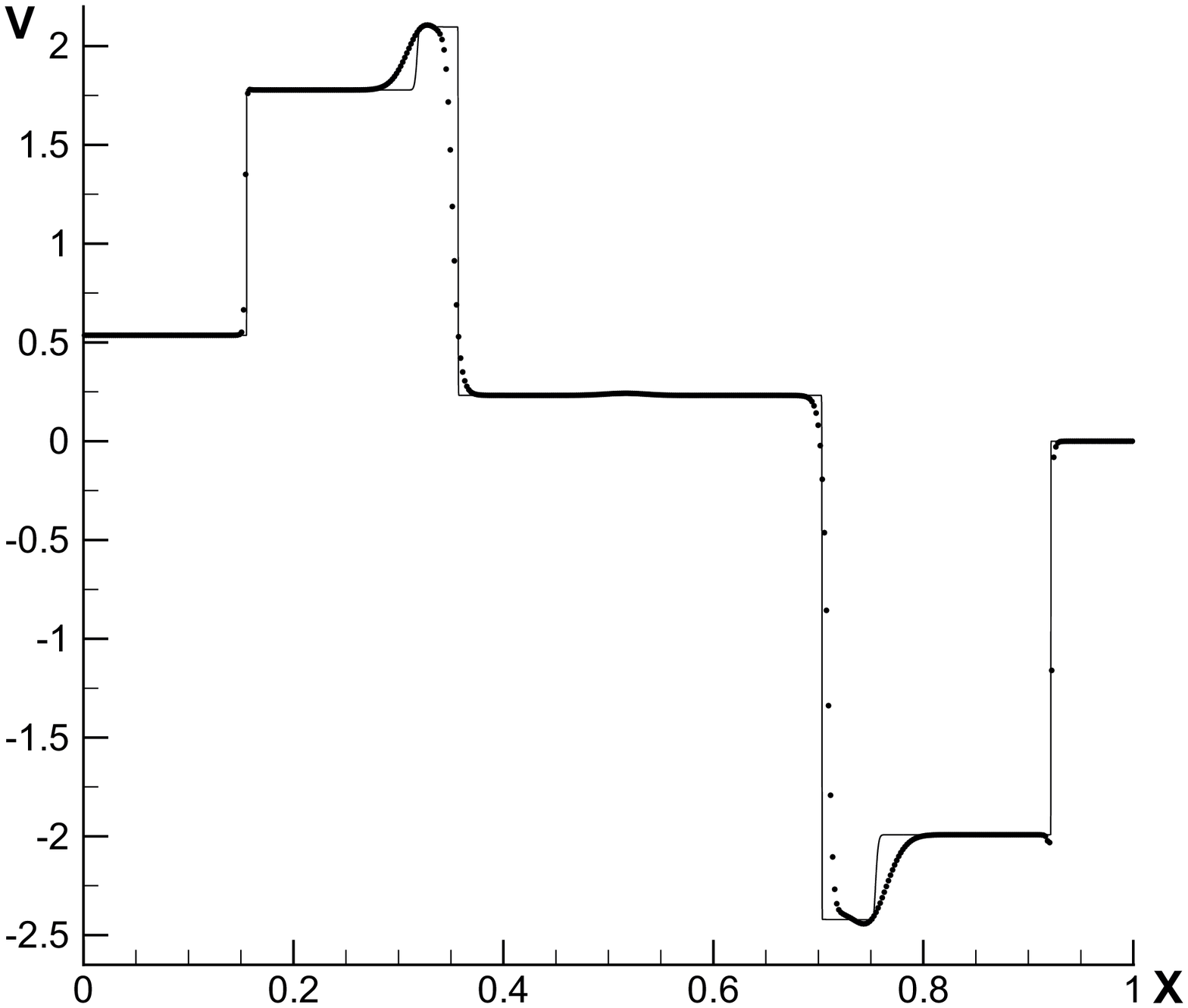}
\caption{Test 5.2. The same for $y$-component of the velocity.
\label{Test3_Vy}}
\end{figure}

\begin{figure}[!h]
\centering
\includegraphics[width=0.48\textwidth]{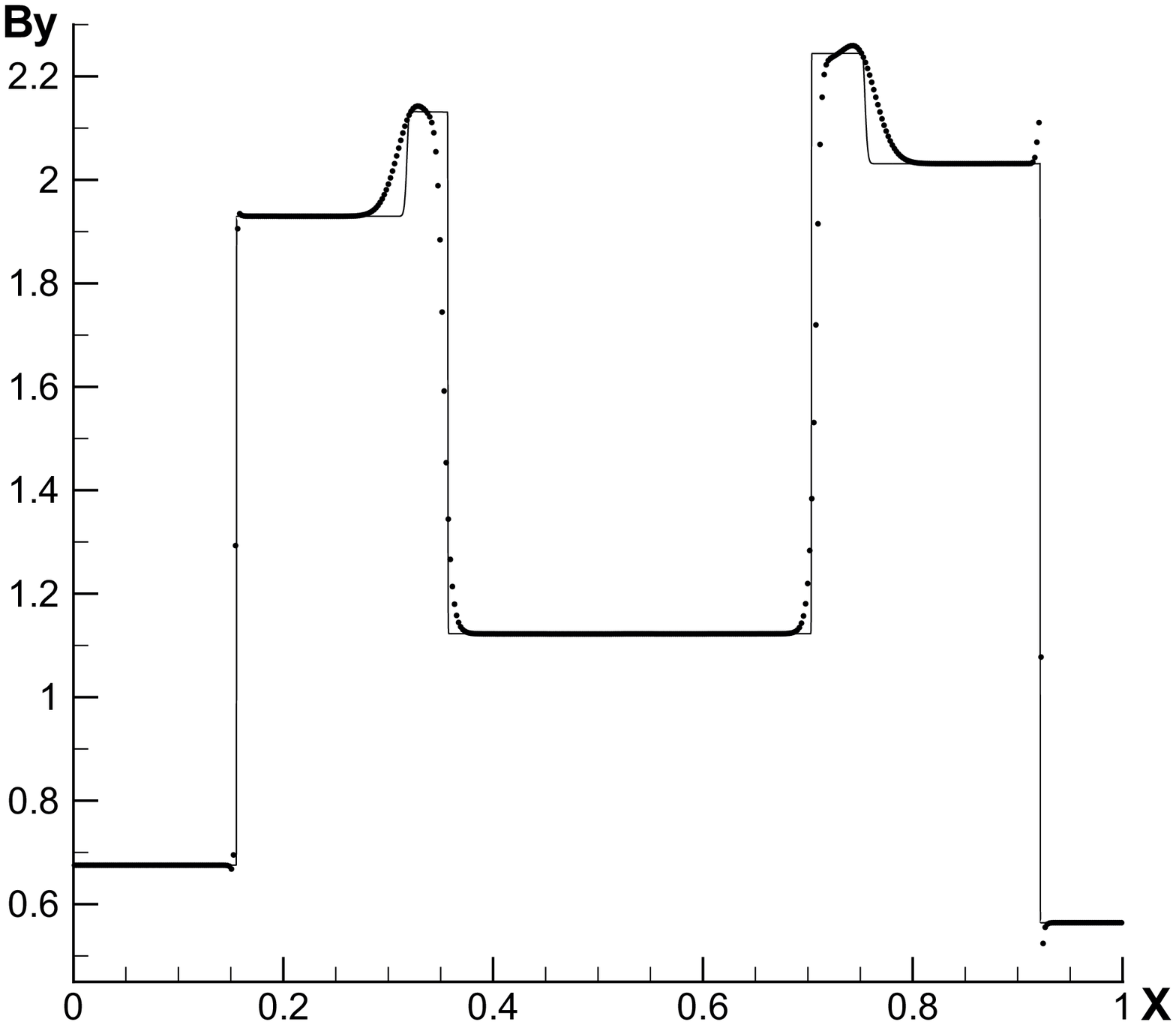}
\caption{Test 5.2. The same for $y$-component of the magnetic field.
\label{Test3_By}}
\end{figure}

Component of the magnetic field $B_x=4/\sqrt{4\pi}$, adiabatic index $\gamma=5/3$, $N=512, T=0.15$, Courant number $\beta = 0.2$, regularization coefficient $\alpha = 0.5$. Computation results are presented on Fig.~\ref{Test3_D}-\ref{Test3_By}.
\begin{figure}[!h]
\centering
\includegraphics[width=0.48\textwidth]{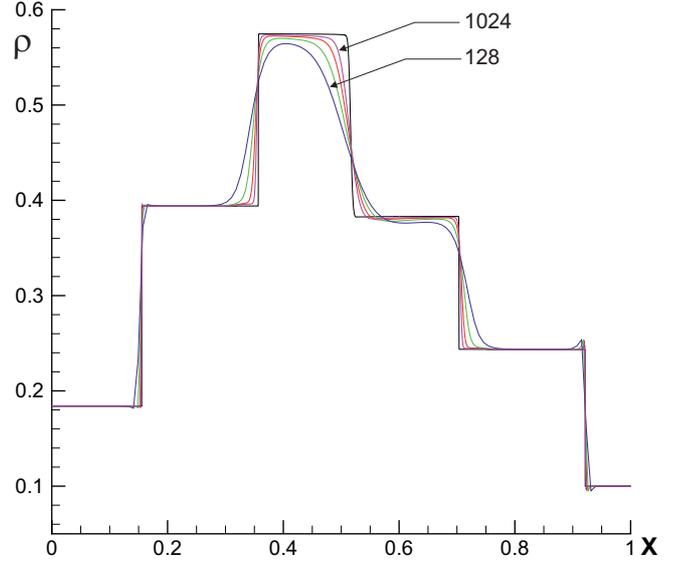}
\caption{Test 5.2. Mesh convergence by example of the density. The computations were performed on the meshes from $N=128$ to $N=1024$ cells.
\label{Test3_D_conv}}
\end{figure}
\begin{figure}[!h]
\centering
\includegraphics[width=0.48\textwidth]{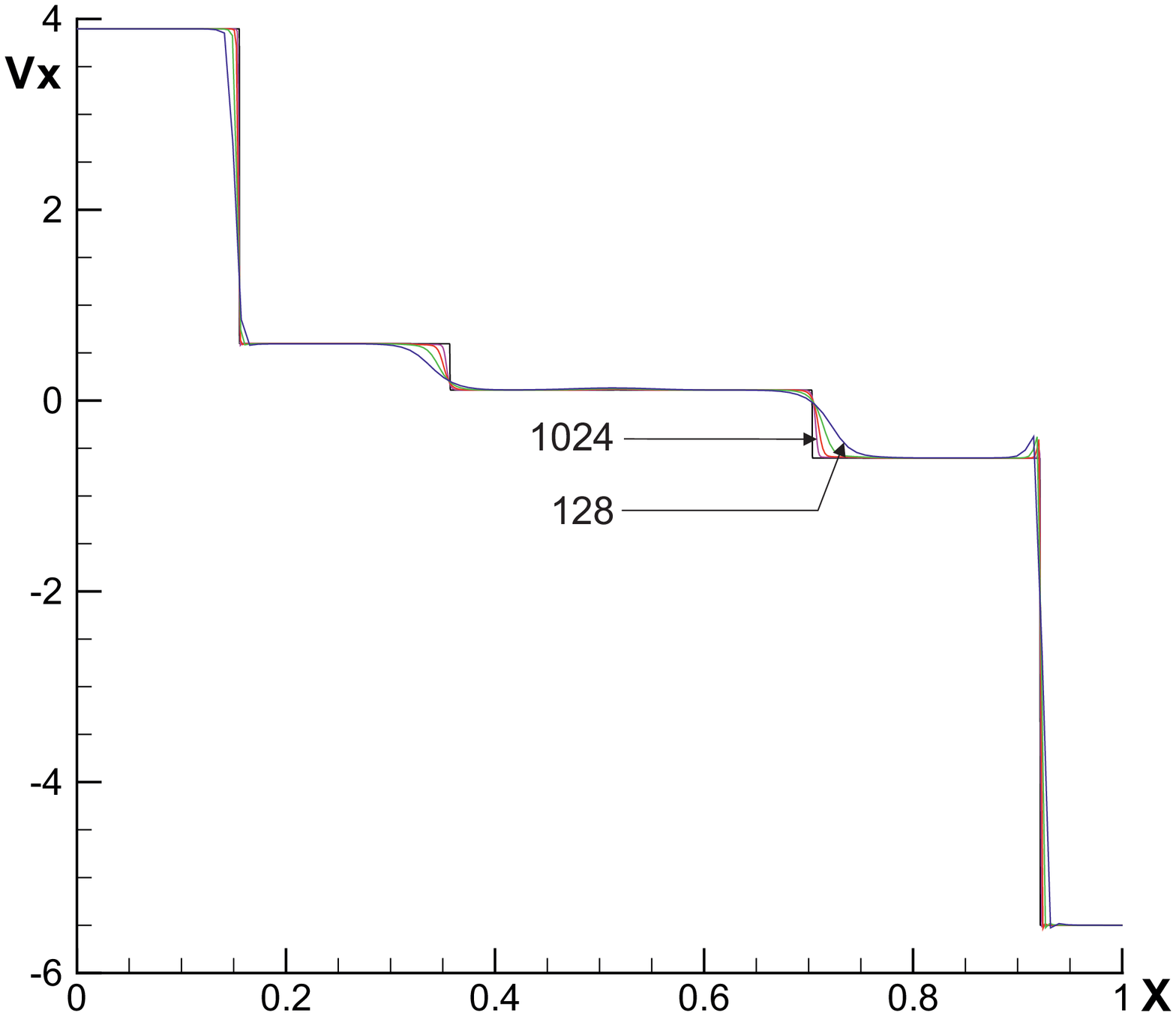}
\caption{Test 5.2. The same for $x$ -component of the velocity.
\label{Test3_Vx_conv}}
\end{figure}
\begin{table}[!h]
\caption{Test 5.2. The average relative errors and the rates of convergence at $T=0.15$.}
\begin{center} \footnotesize
\begin{tabular}{|c|c|c|}
\hline
\rule{0pt}{12pt}\rule{6pt}{0pt}$N$\rule{6pt}{0pt} &\rule{25pt}{0pt} $\delta_N$\rule{25pt}{0pt} &\rule{8pt}{0pt} $R_N$\rule{8pt}{0pt}\\
\hline
128  & $6.47\times10^{-2}$ & -    \\
256 & $3.65\times10^{-2}$ & 0.83\\
512 & $2.05\times10^{-2}$ & 0.84\\
1024 & $1.09\times10^{-2}$ & 0.91\\
\hline
\end{tabular}
\end{center}
\label{table2}
\end{table}

A solution convergence by consecutive reducing of mesh step by two times to exact solution of the problem for the density and the velocity profiles are shown on Fig.~\ref{Test3_D_conv}-\ref{Test3_Vx_conv}. Since the scheme has the first order of accuracy, it is implied from the computations that accurate enough distribution of the density is achieved on fine meshes, while the velocity and the pressure profiles are well resolved on rough meshes (Table~\ref{table2}).

\subsection{Propagation of MHD waves in two-dimensional case}

The next problem is a perfect quantitative test to determine accuracy and convergence order
of a numerical algorithm. All physical parameters in entire computational domain are equal
to constant values to be chosen that main waves are well enough distinct and the wave
vector is directed at some angle to magnetic field. The waves are specified as perturbations to initial constant values of physical quantities in the following form
\begin{equation}
\nonumber
\label{eqxx9}
\delta {\bf U} = A {\bf R} \sin \left(2\pi x\right).
\end{equation}
Here, $\bf U$ is the vector of conservative variables, $A$ is a magnitude, $\bf R$ is a vector of right eigenvectors of hyperbolic MHD system matrix with given numerical values for every wave. In all cases, the magnitude is $A=10^{-6}$. The size of computation domain is equal to the one wave length. Periodic boundary conditions are used for all variables. An error of numerical solution is measured by norm estimation after the wave once passes computational domain
\begin{equation}
\nonumber
\label{eqxx9}
||\delta {\bf U}|| =\sqrt {\sum^{k=N}_{k=1} \left(\delta U_k\right)^2},\quad \delta U_k = \sum_{i} |U^n_{k,i}-U^0_{k,i}|/N,
\end{equation}
Here, $U^n_{k,i}$ is a numerical solution for $k$-th component of vector of conservative
variables for each point $i$ at the time moment $n$, $U^0_{k,i}$ is an initial solution and $N$ represents the number of points in the domain. Initial conditions are given by~\citet{Stur,Gardiner}:
\begin{equation}
\nonumber
\label{eqxx9}
\rho =1, p=\frac {1}{\gamma}, b_x=1, b_y=\sqrt {2}, b_z=0.5,
\end{equation}
where $b=B/\sqrt {4\pi }$ and $\gamma=5/3$.

The values of right eigenvectors components equal to:

for moving to the left fast magnetosonic wave
\begin{multline*}
{\bf R} = (0.4472135954999580, -0.8944271909999160,\\ 0.4216370213557840,
0.1490711984999860,\\ 2.012457825664615, 0.8432740427115680,\\
0.2981423969999720),
\end{multline*}

for moving to the left Alfven wave
\begin{multline*}
{\bf R} = (0, 0,-0.3333333333333333,\\
0.9428090415820634, 0, -0.3333333333333333,\\
0.9428090415820634),
\end{multline*}

for moving to the left slow magnetosonic wave
\begin{multline*}
{\bf R} = (0.8944271909999159, -0.4472135954999579,\\-0.8432740427115680,
-0.2981423969999720,\\  0.6708136850795449, -0.4216370213557841,\\
-0.1490711984999860),
\end{multline*}
where components of the right eigenvectors correspond to the ordering of the conservative
parameters vector $\bf U$ of the form ${\bf U} =[\rho,v_x,v_y,v_z,E,B_y,B_z]$.
Speed of the fast magnetosonic wave is equal to 2, the speed of Alfven wave is equal to 1 and the speed of slow magnetosonic wave is equal to 0.5. The absolute error in propagation of each of these waves and the order of numerical algorithm accuracy are presented in the following tables: for the fast magnetosonic wave (Table~\ref{table_3a}), for the Alfven wave (Table~\ref{table_3b}), for the slow magnetosonic wave (Table~\ref{table_3c}).
\begin{table}[!h]
\caption{Test 5.3. The average relative errors and the rates of convergence at $T=0.15$ for the fast magnetosonic wave.}
\begin{center} \footnotesize
\begin{tabular}{|c|c|c|}
\hline
\rule{0pt}{12pt}\rule{6pt}{0pt}$N$\rule{6pt}{0pt} &\rule{25pt}{0pt} $\delta_N$\rule{25pt}{0pt} &\rule{8pt}{0pt} $R_N$\rule{8pt}{0pt}\\
\hline
  64  & $1.5395\times10^{-7}$ & -    \\
 128 & $8.1368\times10^{-8}$ & 0.9231\\
 256 & $4.1871\times10^{-8}$ & 0.9618\\
 512 & $2.1243\times10^{-8}$ & 0.9823\\
1024  & $1.0700\times10^{-8}$ & 0.9928\\
2048 & $5.3696\times10^{-9}$ & 0.9981\\
\hline
\end{tabular}
\end{center}
\label{table_3a}
\end{table}
\begin{table}[!h]
\caption{Test 5.3. The average relative errors and the rates of convergence at $T=0.15$ for the Alfven wave.}
\begin{center} \footnotesize
\begin{tabular}{|c|c|c|}
\hline
\rule{0pt}{12pt}\rule{6pt}{0pt}$N$\rule{6pt}{0pt} &\rule{25pt}{0pt} $\delta_N$\rule{25pt}{0pt} &\rule{8pt}{0pt} $R_N$\rule{8pt}{0pt}\\
\hline
  64  & $5.6148\times10^{-8}$ & -    \\
 128 & $2.9196\times10^{-8}$ & 0.9467\\
 256 & $1.4920\times10^{-8}$ & 0.9718\\
 512 & $7.5461\times10^{-9}$ & 0.9868\\
1024  & $3.7953\times10^{-9}$ & 0.9949\\
2048 & $1.9033\times10^{-9}$ & 0.9991\\
\hline
\end{tabular}
\end{center}
\label{table_3b}
\end{table}
\begin{table}[!h]
\caption{Test 5.3. The average relative errors and the rates of convergence at $T=0.15$ for the slow magnetosonic wave.}
\begin{center} \footnotesize
\begin{tabular}{|c|c|c|}
\hline
\rule{0pt}{12pt}\rule{6pt}{0pt}$N$\rule{6pt}{0pt} &\rule{25pt}{0pt} $\delta_N$\rule{25pt}{0pt} &\rule{8pt}{0pt} $R_N$\rule{8pt}{0pt}\\
\hline
  64  & $1.2508\times10^{-7}$ & -    \\
 128 & $6.6601\times10^{-8}$ & 0.9124\\
 256 & $3.4399\times10^{-8}$ & 0.9564\\
 512 & $1.7485\times10^{-8}$ & 0.9796\\
1024  & $8.8157\times10^{-9}$ & 0.9914\\
2048 & $4.4262\times10^{-9}$ & 0.9974\\
\hline
\end{tabular}
\end{center}
\label{table_3c}
\end{table}

\subsection{Numerical dissipation and decay of Alfven waves}

In numerical modeling using a space mesh, any numerical scheme always has some dissipation. In order to estimate a level of numerical dissipation of the QMHD scheme a test on decay of Alfven waves was conducted~\citep{Balsara}. At the initial moment of time, the Alfven wave has the following parameters
\begin{equation}
\nonumber
\label{eqxx9}
\delta u_x = u_{amp} c_a \sin \left(k_x x + k_y y\right),
\end{equation}
and moves on fixed background with $\rho_0=1,p_0=1,B_x=1,B_x=B_y=0$.
\begin{figure*}[t]\centering
\parbox[b]{0.32\textwidth}{\centering
\psfig{file=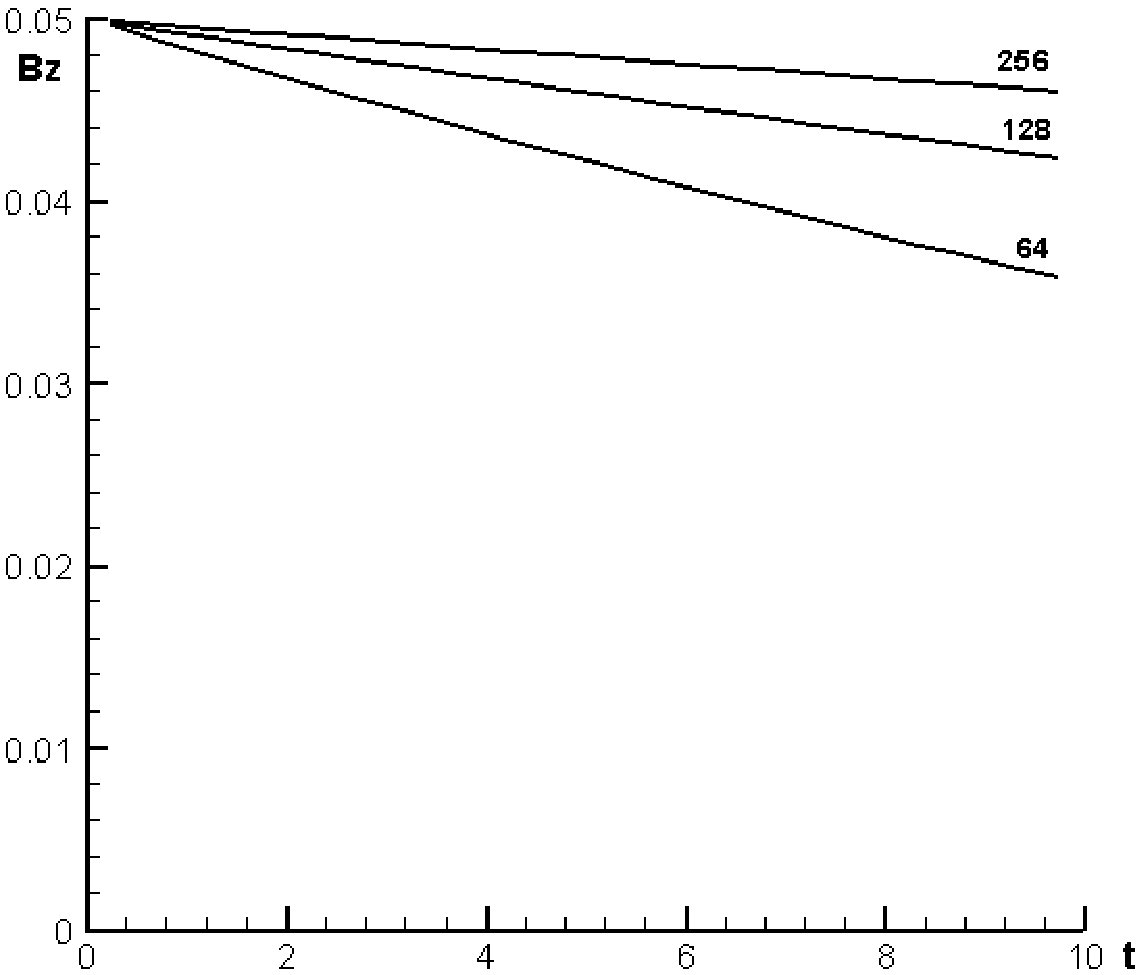,width=0.32\textwidth}
\caption{Test 5.4: Time evolution of the magnetic field $z$-component in computations on various meshes. Values of $N$ are represented by numbers.\label{4D_1}}}
\hfil
\begin{minipage}[b]{0.32\textwidth}
\centering
\psfig{file=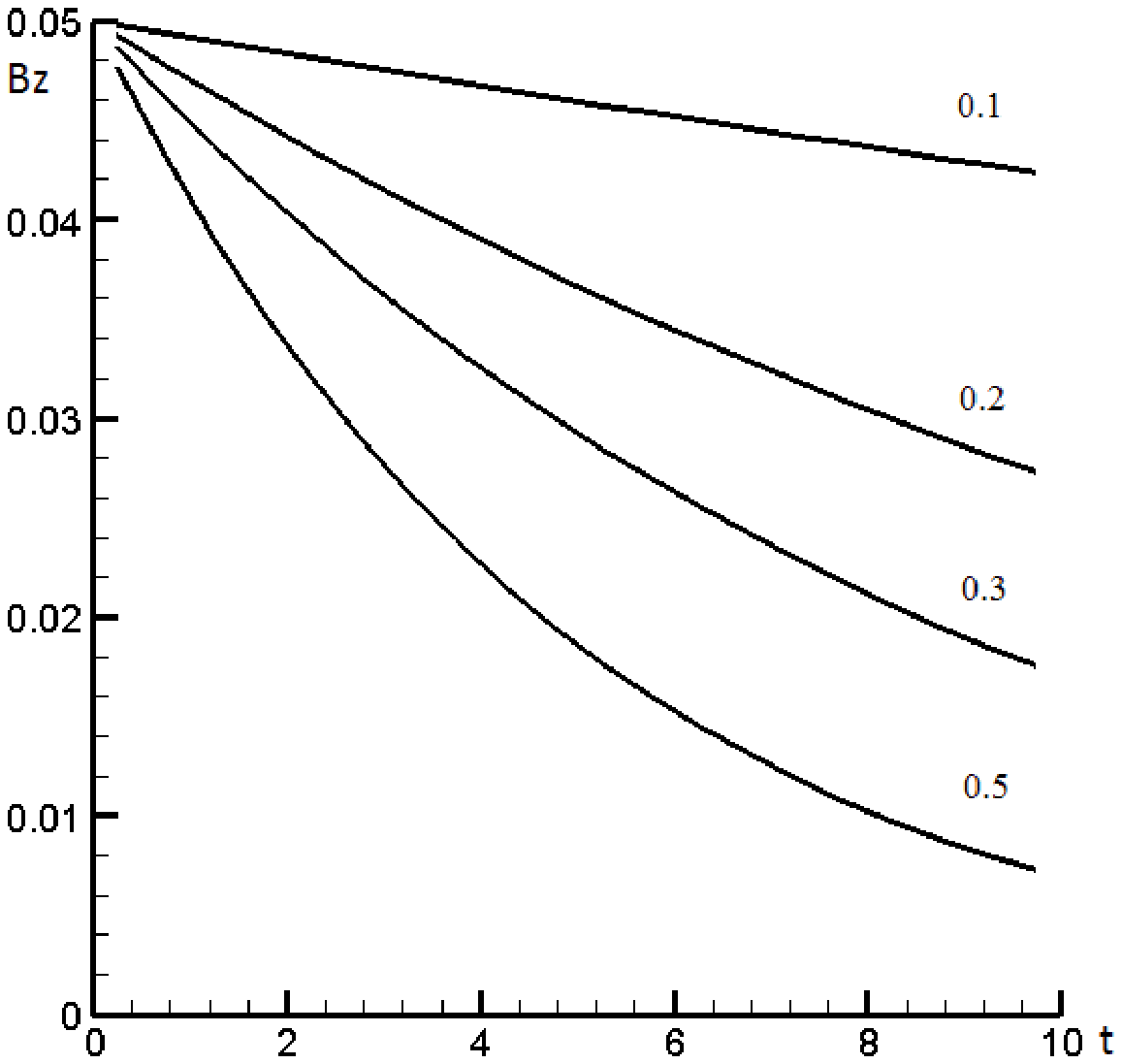,width=0.96\textwidth}
\caption{Test 5.4: The same on the mesh with $N=128$ at various values $\alpha$.\vspace{1.0em}\label{4D_2a}}
\end{minipage}
\hfil
\begin{minipage}[b]{0.32\textwidth}
\centering
\psfig{file=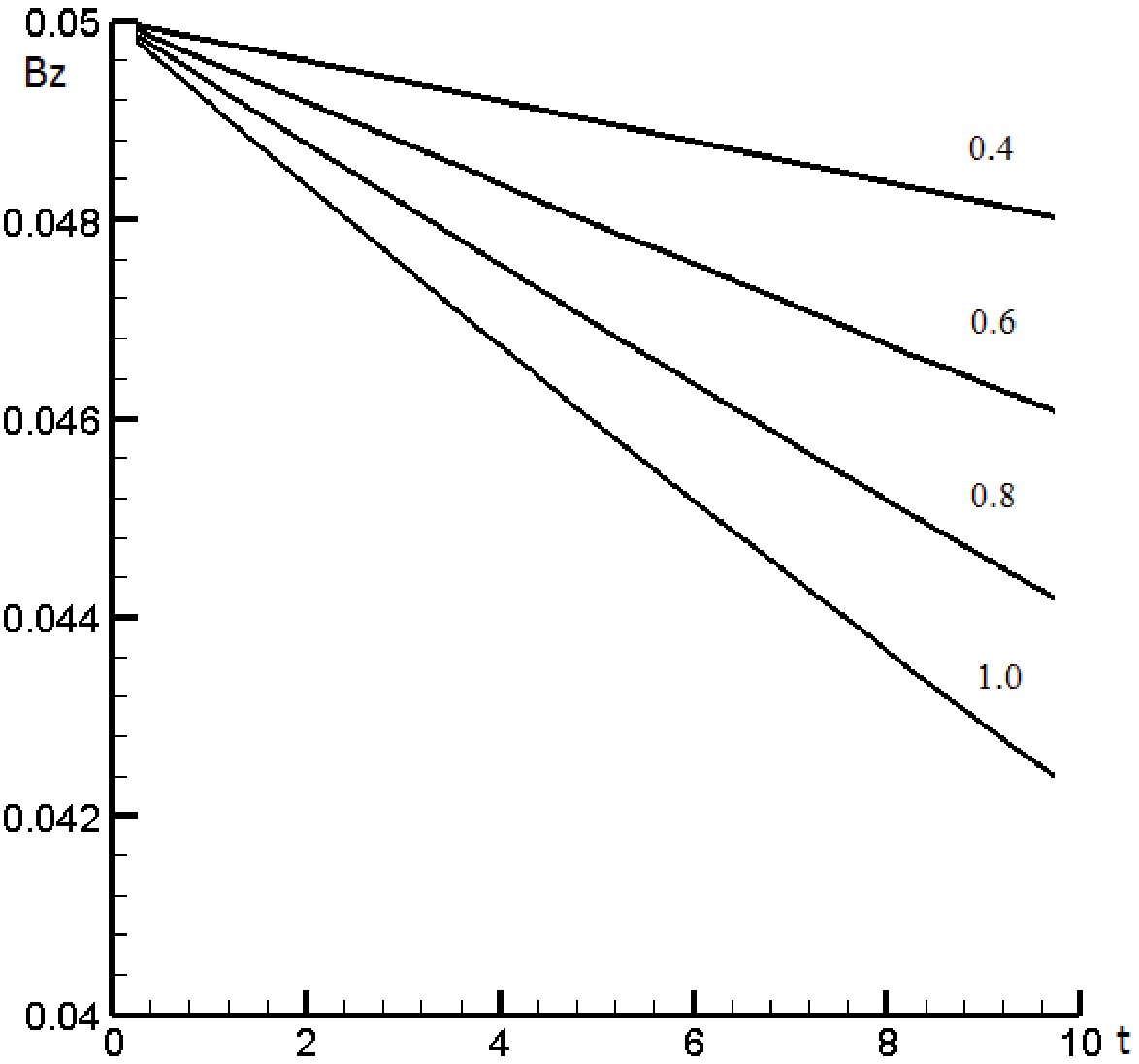,width=0.96\textwidth}
\caption{Test 5.4: The same at various Schmidt numbers $Sc$.\vspace{1.0em}\label{4D_2b}}
\end{minipage}
\end{figure*}

Computational domain represents a square of side $L=1$. Computations were performed on three meshes with number of cells in each direction $N=64$, $128$, $256$. The initial Alfven wave speed is $c_a=0.7071$ and its magnitude is $u_{amp}=0.1$, the adiabatic index is $\gamma=5/3$. Computations were done with the Courant number $\beta=0.3$, parameter $\alpha =0.1$ and with implementation of the periodic boundary conditions. Fig.~\ref{4D_1} shows the time evolution of magnetic field $z$-component maximum on a sequence of twice refined meshes. Dissipation level corresponds to the schemes of the first order by space and time, and quickly decreases with mesh refinement. With the given Courant number and number of the mesh cells, Figs.~\ref{4D_2a}-\ref{4D_2b} show that dissipation of the numerical scheme decreases with $\alpha$ and $S\!c$ parameters reducing.

The smallest dissipation level of the numerical scheme, wherein solution stability preserves, corresponds to values of parameters $\alpha =0.1$ and Schmidt number $S\!c=0.4$ on the mesh with $N =128$. In this case computational results are similar to results obtained with PPML method for $N=64$~\citep{Pop08,Ust09}.

\subsection{Propagation of a circularly polarized Alfven wave}

This test problem was considered in the paper~\citet{Gardiner} to study the accuracy and the order of
convergence of numerical schemes on smooth solutions. The Alfven wave propagates along diagonal
of the mesh at the angle $\theta=\tan^{-1}(0.5)\approx 26.6^{{\rm o}}$ to axis $x$. Computational
domain has a size of $L_x=2L_y$, with cells number of $N_x=2N_y$. Since the wave does not move
along diagonals of discrete cells, the problem has real multidimensional nature.  Initial conditions
are given by:
\begin{align*}
\nonumber
\label{eqxx9}
& \rho =1, u_{||}=0, u_{\bot}=0.1\sin(2\pi\xi), u_z=0.1\cos(2\pi\xi),\\
& p =1, B_{||}=0, B_{\bot}=0.1\sin(2\pi\xi), B_z=0.1\cos(2\pi\xi),
\end{align*}
where $\xi=x\cos\theta+y\cos\theta$. Here, $u_{||},u_{\bot},B_{||},B_{\bot}$ are components
of the velocity and the magnetic field, directed in parallel and perpendicular to the direction
of the Alfven wave movement. The wave propagates towards a point $(x,y)=(0,0)$ with speed $B_{||}/\sqrt \rho=1$.
The problem was solved with numerical cells of $N=16,32,64,128,256$ in the direction $x$,
herewith relative numerical error was estimated for each quantity by the formula
\begin{equation}
\nonumber
\label{eqxx9}
\delta_N(U) = \frac {\sum^{i=2N}_{i=1} \sum^{i=2N}_{j=1} |U^N_{i,j}-U^E_{i,j}|} {\sum^{i=2N}_{i=1} \sum^{i=2N}_{j=1} |U^E_{i,j}|}\\
,\quad U = u_{\bot}, u_z, B_{\bot}, B_z,
\end{equation}
where $U^E_{i,j}$ is the exact solution. Convergence order of the scheme was estimated as
\begin{equation}
\nonumber
\label{eqxx9}
R_N = \log_2 \left(\delta_{N/2}/\delta_N\right),
\end{equation}
where $\delta_N$ was defined as mean by
\begin{equation}
\nonumber
\label{eqxx9}
\delta_N = \frac {1}{4} \left(\delta_N(u_{\bot})+\delta_N(u_z)+\delta_N(B_{\bot})+\delta_N(B_z)\right),
\end{equation}

\begin{figure}[!t]
\centering
\includegraphics[width=0.48\textwidth]{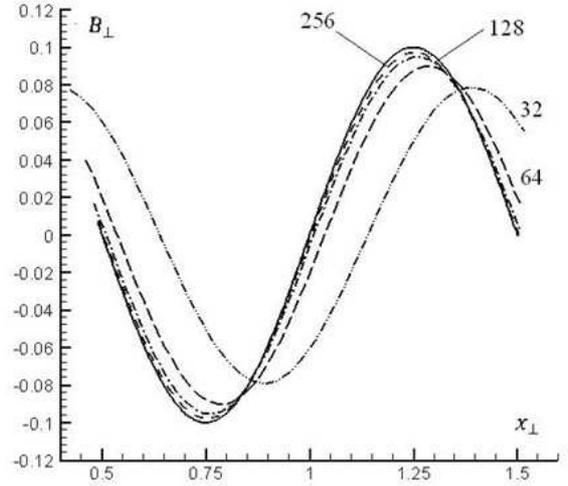}
\caption{Test 5.5: The lines of orthogonal component of the magnetic field in the computations
on various meshes. The values of N are indicated by numbers.
\label{5D_1}}
\end{figure}

The computations were carried out until the time moment of $t=5$ with the Courant number of $\beta=0.2$ , the parameter $\alpha =0.1$, the adiabatic index $\gamma=5/3$, and with the periodic boundary conditions were used. On the Fig.~\ref{5D_1}, orthogonal component $B_{\bot}$ of the magnetic field is presented in computations on various meshes. The values of $N$ are indicated by numbers. It can be seen that the numerical solution of the problem tends to the exact solution with increasing of $N$. Obtained results confirm that with increasing of the resolution the numerical scheme has the first order of accuracy in space and time. The same test was performed by PPML code in~\citet{Pop08}. For PPML method acceptable numerical solution begins from the grid resolution equal to 32 cells, but for QMHD -- from 64 cells. 

\begin{figure}[!h]
\centering
\includegraphics[width=0.48\textwidth]{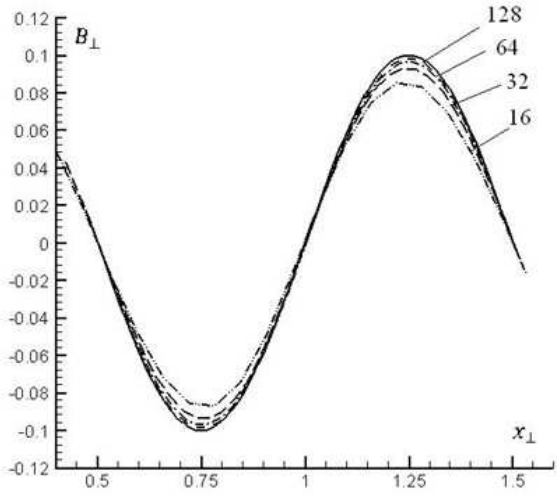}
\caption{Test 5.5: The lines of the perpendicular component of the magnetic field in
the computations on various meshes in the case of the stationary wave. The values of $N$ are
indicated by numbers.
\label{5D_2}}
\end{figure}

Also, computations were performed for the case of the stationary Alfven wave with $u_{||}=1$. On Fig.~\ref{5D_2}, the orthogonal component $B_{\bot}$ of the magnetic field are presented in computations on various meshes. The values of $N$ are indicated by numbers. In Table~\ref{table_56}, the average relative error and the order of convergence are shown. It can be seen that the numerical solution of the problem with increasing of $N$ rapidly tends to the exact solution and the order of convergence is close to one even for small $N$. These results are compared with high-order Constrained Transport/Central Difference(CT/CD) scheme~\citep{Toth}, see Table~\ref{table_57}. Comparison of the tables shows, that  for $N=16$ the accuracy of both methods are similar, and for $N=32$ and 64 the accuracy of Flux-CD/CT scheme is approximately two times higher.

\begin{table}[!h]
\caption{Test 5.5. The average relative errors and the rates of convergence with $u_{||}=1$}
\begin{center} \footnotesize
\begin{tabular}{|c|c|c|}
\hline
\rule{0pt}{12pt}\rule{6pt}{0pt}$N$\rule{6pt}{0pt} &\rule{25pt}{0pt} $\delta_N$\rule{25pt}{0pt} &\rule{8pt}{0pt} $R_N$\rule{8pt}{0pt}\\
\hline
 16  & 0.12671 & -    \\
 32 & 0.064888 & 0.9688\\
 64 & 0.032914 & 0.9825\\
 128 & 0.016569 & 0.9935\\
 256 & 0.0083133 & 0.9984\\
\hline
\end{tabular}
\end{center}
\label{table_56}
\end{table}

\begin{table}[!h]
\caption{Test 5.5. The same for $u_{||}=1$ by Flux-CD/CT scheme taken from~\citet{Toth}.}
\begin{center} \footnotesize
\begin{tabular}{|c|c|c|}
\hline
\rule{0pt}{12pt}\rule{6pt}{0pt}$N$\rule{6pt}{0pt} &\rule{25pt}{0pt} $\delta_N$\rule{25pt}{0pt} &\rule{8pt}{0pt} $R_N$\rule{8pt}{0pt}\\
\hline
 8  & $0.315$ & -    \\
 16  & $0.122$ & 1.368\\
 32 & $0.037$ & 1.721\\
 64 & $0.013$ & 1.509\\
\hline
\end{tabular}
\end{center}
\label{table_57}
\end{table}

\subsection{A blast wave propagation through magnetized medium}

\begin{figure}[!t]
\centering
\includegraphics[width=0.48\textwidth]{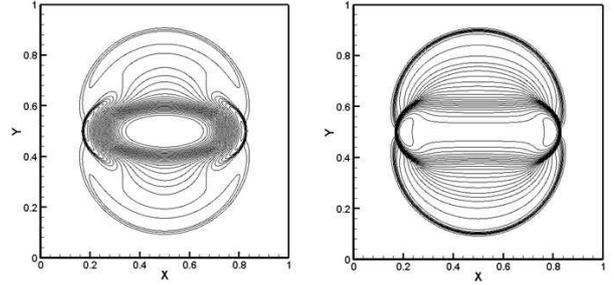}
\caption{Test 5.6: Distribution of logarithm of the density and logarithm of the pressure.
\label{6D_1}}
\end{figure}

\begin{figure}[!t]
\centering
\includegraphics[width=0.48\textwidth]{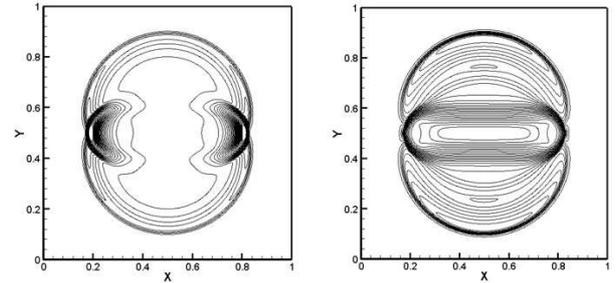}
\caption{Test 5.6: Distribution of the kinetic energy and logarithm of the magnetic energy.
\label{6D_2}}
\end{figure}

In this problem, a propagation of the initial finite perturbation of the pressure through a medium with superimposed magnetic field~\citep{Woodward} is investigated. The problem is solved in the square computational domain with side dimension  $L=1$ and number of cells $400\times400$. In the initial time, in the entire domain the initial density $\rho=1$ and pressure $p=1$, excepting central part with radius $r=0.05$, where pressure is  $p=1000$. Uniform magnetic field with magnitude  $B=10$  is directed along axis $x$. The adiabatic index $\gamma=1.4$. Computations were carried out until time $t=0.02$ with the Courant number $\beta=0.1$, the parameter $\alpha =0.4$. Gradients of all parameters were equal to zero on the domain boundary. Figs.~\ref{6D_1}-\ref{6D_2} present the numerical solution at $t=0.02$.


Magnetic field introduces anisotropy in a substance expansion. Under the action of the pressure,
substance accelerates along magnetic field lines, and shock waves with higher kinetic and magnetic
energy can be seen on the boundary in the middle part of the domain. A rarefied area with lower
density and pressure and the prevalence of magnetic energy over the kinetic and thermal energy
forms in the center of the square. In spite of the large initial difference in the pressure and
high magnetization of the medium in the middle part of the computation domain, the numerical
QMHD scheme provides positive values of the pressure and density, and describes all intrinsic
discontinuities with good accuracy for the scheme of the first order in space and in time at the
final stage of substance expansion.

\subsection{Two-dimensional Riemann problem with four states with the magnetic field}

Structures formation is studied in the interaction of the four states with a superimposed
magnetic field~\citep{Woodward,Arminjon}. Initial conditions are given by:
\begin{align*}
\nonumber
(\rho, p, u_x, u_y) &= (1, 1, 0.75, 0.5),&x>0, y>0\\
(\rho, p, u_x, u_y) &= (2, 1, 0.75, 0.5),&x<0, y>0\\
(\rho, p, u_x, u_y) &= (1, 1, -0.75, 0.5),&x<0, y<0\\
(\rho, p, u_x, u_y) &= (3, 1, -0.75, -0.5),&x>0, y<0
\end{align*}
\begin{figure}[!t]
\centering
\includegraphics[width=0.48\textwidth]{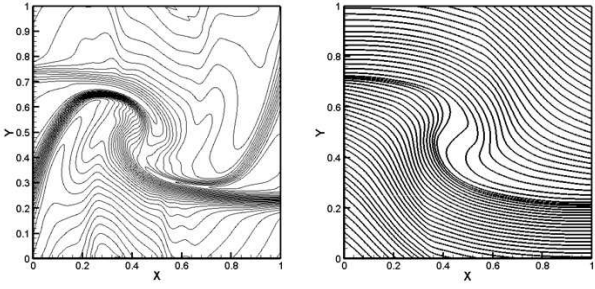}
\caption{Test 5.7: Distributions of density contour lines and magnetic field lines.
\label{7D_1}}
\end{figure}
\begin{figure}[!t]
\centering
\includegraphics[width=0.48\textwidth]{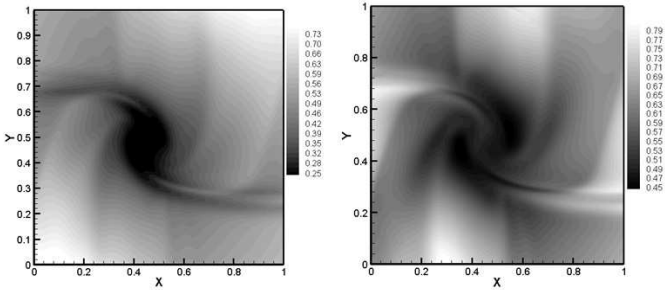}
\caption{Test 5.7: Distributions of the magnetic energy and pressure.
\label{7D_2}}
\end{figure}

The problem is solved in the square domain with side dimension  $L=1$. The uniform magnetic field ${\bf B}=(2,0,1)$ is imposed. A numerical solution is computed up to time of $t=0.8$ on the mesh of $400\times400$ with the Courant number of $\beta=0.1$, the parameter $\alpha =0.4$, the adiabatic index $\gamma=1.4$. Gradients of
all parameters were equal to zero on the domain boundary. Figs.~\ref{7D_1}-\ref{7D_2} show results of the numerical solution at $t=0.8$. In the center of the square, a rarefied area forms with lower pressure and magnetic energy, the magnetic field lines are strongly curved and diverge. In the middle of the square far from the center on the edges of the vortex area, formation of the compression waves can be seen with increasing of the density, pressure, magnetic energy and a large concentration of the magnetic field lines. Contour lines of the density and magnetic energy describe the structure of the vortex flow with enough precision.

\subsection{Orszag-Tang Vortex}

\begin{figure}[!t]
\centering
\includegraphics[width=0.48\textwidth]{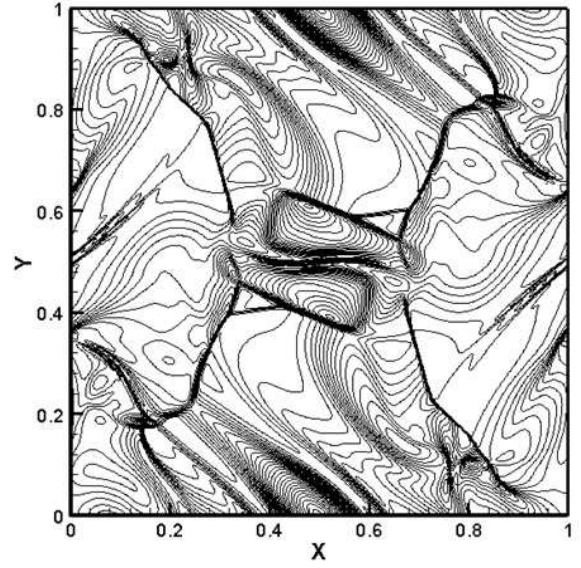}
\caption{Test 5.8: Distribution of the pressure contour lines in the range of $0.05$ to $0.5$ with step $0.015$
at the moment $t=0.5$.
\label{8D_1}}
\end{figure}

In this problem, the formation of the complex structure of shock waves in supersonic turbulence
is considered~\citep{Arminjon}. The problem is solved in square domain with side $L=1$. Initial conditions are given by:
\begin{gather*}
\rho =25/(36\pi), p=5/(12\pi), u_x=-\sin (2\pi y), u_y=\sin (2\pi x),\notag \\ u_z=0,
B_x=-B_0\sin (2\pi y), B_y=B_0\sin (4\pi x), B_z=0,
\end{gather*}
A solution is computed up to time $t=2$ on the meshes $400\times400$    and $800\times800$ with the Courant number of $\beta=0.2$ and the parameter $\alpha =0.3$, the adiabatic index $\gamma=5/3$. Periodic boundary conditions were used. Fig.~\ref{8D_1} shows distribution of the pressure contour lines at time $t=0.5$.
\begin{figure}[!h]
\centering
\includegraphics[width=0.48\textwidth]{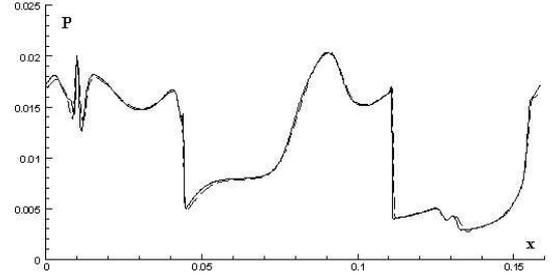}
\caption{Test 5.8: Pressure along line $y=0.3125$.
\label{8D_2}}
\end{figure}
\begin{figure}[!h]
\centering
\includegraphics[width=0.48\textwidth]{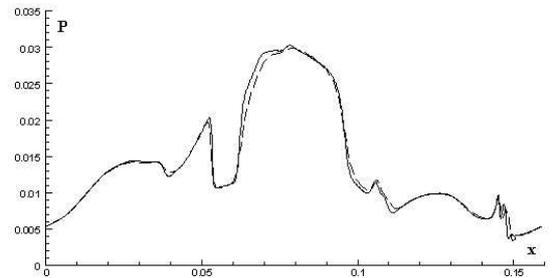}
\caption{Test 5.8: Pressure along line $y=0.4277$.
\label{8D_3}}
\end{figure}

\begin{figure}[!t]
\centering
\includegraphics[width=0.48\textwidth]{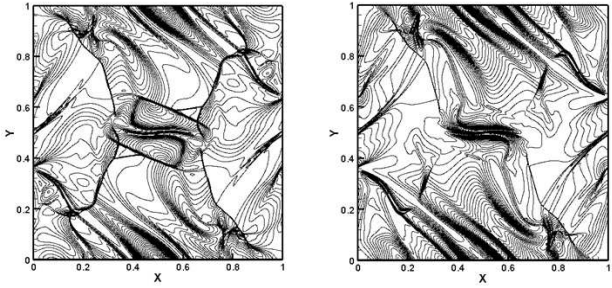}
\caption{Test 5.8: Distributions pressure and magnetic energy on the mesh $800\times800$
\label{8D_4}}
\end{figure}

Figs.~\ref{8D_2}-\ref{8D_3} show distributions of pressure in the two-dimensional plane sections along the lines $y=0.3125 (j=130)$ and $y=0.4277(j=176)$, allowing to estimate a reproducing accuracy of the discontinuities in the solution. Result on mesh with $N=400$ is depicted by dash line and solid line for $N=800$. Fig.~\ref{8D_4} shows distribution of the  pressure and magnetic energy contour lines for $t=0.5$ on the mesh of $800\times800$. Comparing to the results, obtained with the high-order PPML method~\citep{Ust09}, we can say that QMHD gives the correct structure of the flow with all discontinuities. On the fine grid the accuracy of QMHD is acceptable.

\subsection{Interaction of a shock wave with a cloud}

The problem of a cloud destruction by a shock wave in a magnetic field~\citep{Toth} is considered here. The problem is solved in a square domain with a side $L=1$. In the initial moment of time, two stationary states are defined
\begin{gather*}
\nonumber
U_l =(3.86859,11.2536,0,0,167.345,\\ 0,2.1826182,-2.1826182),\\
U_r =(1, 0, 0, 0, 1, 0, 0.56418958, 0.56418958),
\end{gather*}
related to the shock wave, and separated by a plane $x=0.05$.
\begin{figure}[!t]
\centering
\includegraphics[width=0.48\textwidth]{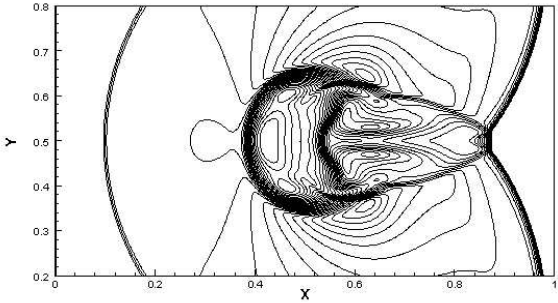}
\caption{Test 5.9: The contours represent levels of logarithm of the density.
\label{9D_1}}
\end{figure}
\begin{figure}[!h]
\centering
\includegraphics[width=0.48\textwidth]{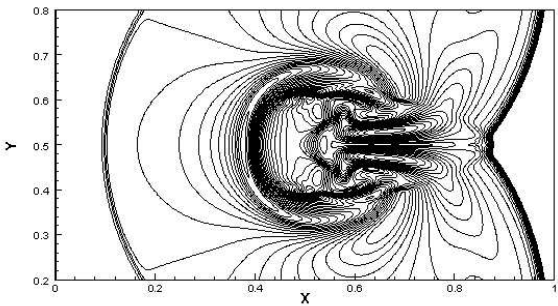}
\caption{Test 5.9: The same for logarithm of the magnetic energy.
\label{9D_2}}
\end{figure}
\begin{figure}[!h]
\centering
\includegraphics[width=0.48\textwidth]{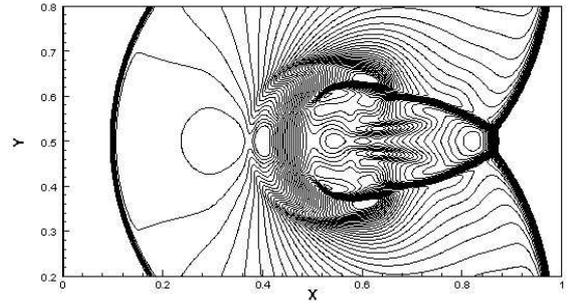}
\caption{Test 5.9: The same for logarithm of the pressure.
\label{9D_3}}
\end{figure}

A spherical cloud with  density $\rho=10$ and radius $r=0.15$ is set in hydrostatic equilibrium
with the environment at the point with coordinates $(0.3,0.5)$. The solution is computed at time moment of $t=0.06$ on the mesh of $400\times400$ with the Courant number $\beta=0.1$, the parameter $\alpha =0.4$. The adiabatic index $\gamma=5/3$.  Gradients of all parameters were equal to zero on the domain boundary. Figs.~\ref{9D_1}-\ref{9D_3} show distributions of contour levels of logarithm of the density, magnetic energy and pressure correspondingly.

\vspace{-6pt}

\section{Conclusions}
\vspace{-2pt}

In this paper, we presented the extension of quasi-gas dynamic approach for a solution of ideal problems of magneto\-hydro\-dynamics. We obtained the regularized, or quasi-gas dynamic (QMHD) system of equations for ideal magneto\-hydro\-dynamics by applying temporal averaging to all physical parameters, including the magnetic field. The numerical QMHD scheme is multidimensional, where evolution of all physical quantities is carried out in unsplit form by space directions.

It is shown that based on a single approach, the QMHD computational method applied for a compressible magneto\-hydro\-dynamics allows modeling a wide range of non-stationary MHD problems.
 For all studied test cases, computation shows steady converging of numerical solution to its exact solution with shredding of space mesh, providing accurate representation of distribution for all physical quantities on smooth part of a solution and on discontinuities as well. With adjustment of the tuning parameters the quality of the numerical solution may be increased.

The tests of Orszag-Tang vortex, a blast wave propagation through magnetized medium, and interaction of a shock wave with a cloud have been  solved recently by QMHD scheme for full 3D case~\citep{Pop13}. The value of tuning parameters $\alpha=0.5$, $Sc=Pr=1$ are suitable for all problems in 1D, 2D and 3D cases.

 Simplicity of numerical realization and uniformity of the algorithm provides natural realization on parallel computer systems implying domain-decomposition technique. The second makes QMHD approach promising for numerical solution of complex  3D MHD problems.

 The disadvantage of the QMHD method is the  first-order  approximation, that requires more detailed computational mesh to obtain the quality of a solution in comparison  to high-order schemes, such as PPML. 
On the other hand, the high approximation orders are a little bit formal since they are true  only for smooth solutions, whereas in most interesting applications the solutions are discontinuous . 

Still QMHD method is robust, relatively cheap by  computational cost and requires no additional monotonization procedures, e.g. limiting functions, what is very nice property especially for magneto\-hydro\-dynamic simulations.

\vspace{-6pt}

\section*{Acknowledgement}
\vspace{-2pt}

This work was accomplished with financial support of Russian Foundation for Basic Research (project na. 12-02-31737-mol\_a, 13-01-00703).

\bibliographystyle{elsarticle-harv}

\end{document}